\documentclass[12pt, a4paper]{article}
\usepackage{amsmath,amssymb,amsfonts,esint}
\usepackage{graphicx,caption,float}
\usepackage[colorlinks]{hyperref}
\usepackage{xcolor} 

\begin{document}

\title{Magnetic flux in the vacuum of quantum bosonic matter in the cosmic string background}

\author{ Yurii A. Sitenko${}^{1,2,3}$, Volodymyr M. Gorkavenko${}^{4}$, and 
Maria S. Tsarenkova${}^{4}$\\
\it \small ${}^{1}$Bogolyubov Institute for Theoretical Physics,
 \it \small National Academy of Sciences of Ukraine,\\
 \it \small 14-b Metrologichna str., Kyiv 03143,
 Ukraine\\\phantom{bhh}\\
 \it \small ${}^{2}$Van Swinderen Institute for Particle Physics and Gravity, University of Groningen,\\ \it \small 4 Nijenborg, Groningen
 9747 AG, Netherlands\\\phantom{bhh}\\
 \it \small ${}^{3}$Erwin Schr\"{o}dinger International Institute for Mathematics and Physics,
 University of Vienna,\\ \it \small 9/2 Boltzmanngassse, Vienna
 A-1090, Austria\\\phantom{bhh}\\
 \it \small ${}^{4}$Department of Physics, Taras Shevchenko National
 University of Kyiv,\\ \it \small 64 Volodymyrs'ka str., Kyiv
 01601, Ukraine}
 \date{}
\maketitle

\begin{abstract}
The relativistic spin-0 matter field is quantized in the background of a straight cosmic string with nonvanishing transverse size. The most general boundary condition ensuring the impenetrability of the matter field into the interior of the cosmic string is shown to be the Robin condition with a  boundary parameter varying arbitrarily from point to point of the boundary. The role of the bound states in the spectrum of solutions to the Fock-Klein-Gordon equation is elucidated. We derive, in the general case, an analytic expression for the total magnetic flux which is induced  in the vacuum in the cosmic string background. The further numerical analysis and requirement of physical plausibility are shown to restrict ambiguity which is due to the boundary condition. The dependence of the induced vacuum magnetic flux on the string flux and tension, as well as on the transverse size of the string, is analyzed.
\end{abstract}

\section{Introduction}

Considerable attention is always paid to the study of nonperturbative effects in quantum systems, arising as a consequence of the interaction of quantized fields with various configurations of classical fields. Especially interesting is the influence of configurations with nontrivial topology (domain walls, vortices, monopoles, or, in general, topological defects) on the properties of quantum systems. In the course of spontaneous breaking of a continuous symmetry, a confined region of the false vacuum of the appropriate Higgs field forms a locus of trapped energy, i.e., a topological defect, which can serve as a background for quantum matter fields. There is a need, in this regard, to take into account the finite size of a topological defect and to set up a boundary condition on its edge. An approach might be to employ a family of the most general boundary conditions ensuring the impenetrability of quantum matter fields into the interior of a topological defect; in mathematical parlance, this means the condition of self-adjointness for the appropriate quantum-mechanical operator of one-particle energy. Then a task is to discover effects that are induced by the topological defect in the ground state of the quantum matter system in the general case, while a further analysis with the requirement of physical plausibility of obtained results is aimed to restrict an arbitrariness in the choice of boundary conditions. In this way, there is a possibility to achieve the unambiguous determination of the influence of the topological defect on quantum matter.

In the present paper, a topological defect in the form of the Abrikosov-Nielsen-Olesen vortex \cite{Abr,NO} is considered. Such defects are known in cosmology and astrophysics under the name of cosmic strings, they emerge in the aftermath of phase transitions with spontaneous gauge symmetry breaking during evolution of the early universe \cite{Ki2,Vil2}. Cosmic strings, starting from a random tangle, evolve into two different sets: the unstable one which consists of a variety of string loops decaying by gravitational radiation and the stable one which consists of several long, approximately straight strings spanning the horizon, see, e.g., reviews in \cite{Vil3,Ki3}. Although observational bounds predict a negligible contribution of cosmic strings to large scale inhomogeneity such as the angular distribution in the cosmic microwave background radiation, their evolution brings distinct astrophysical effects, in particular, they produce detectable gravitational waves \cite{Dam}, gamma-ray bursts \cite{Ber}, and high-energy cosmic rays \cite{Bhat}. The interest in cosmic strings is augmented by theoretical findings that they are demanded in the framework of the  superstring theory inspired cosmological models, i.e., the brane-world models of inflation \cite{Jean,Dva,Pol}. While the universe undergoes phase transitions at energy scales much higher than the scale of grand unification, such supermassive cosmic strings can be produced, opening new observational windows, see \cite{Sak,Cop}.

Vortex-type defects are widely discussed in the context of condensed matter physics as well, in particular, they can be viewed as disclinations in nanoconical structures.  A development in material science of this century provides a remarkable link between condensed matter and high energy physics, which is caused to a large extent by the experimental discovery of graphene -- a two-dimensional crystalline allotrope formed by a monolayer of carbon atoms \cite{Nov,Ge}. It is well established by now that a sheet of graphene is always corrugated and covered by ripples that can be either intrinsic or induced by roughness of a substrate \cite{Kats}. A single topological defect (disclination) warps a sheet of graphene, rolling it into a nanocone, which is similar to the transverse section of a spatial region out of a cosmic string, see, e.g., \cite{Kri,SiV7,Nae,Si18}.

Spin-0 quantized fields in various aspects are a subject of study in cosmology and astrophysics. They are present in all unified field theory models, appearing as possible types of matter, in particular, as dilatons and inflatons in the early universe, as candidates to describe dark matter, and as possible Bose-Einstein condensates. Our purpose is to study the influence of a vortex-type defect on the ground state of the surrounding quantum relativistic bosonic (scalar or pseudoscalar) matter. The induced ground state magnetic flux will be found, and its dependence on the gauge flux and the string tension, as well as on the defect size and the choice of boundary conditions, will be analyzed. We show that the boundary conditions allowing for the existence of bound states are incompatible with the physically plausible behavior of the induced ground state magnetic flux.

In the next section we define the current and the magnetic field strength that are induced by a cosmic string in the vacuum of quantum relativistic bosonic matter. In Section 3 we determine the most general boundary condition ensuring the impossibility for matter to penetrate through the edge of the string core and display an appearance of bound states in addition to continuum ones. The analytic expression for the induced vacuum magnetic flux in the general case is derived in Section 4. In Section 5 the numerical study of the analytic expression is carried out, and we show that the range of the boundary parameter values is restricted by requiring the physically plausible behavior for the flux. Finally, the results are summarized and discussed in Section 6. The details of derivation of the asymptotics of the flux at small values of the transverse size of the string are given in Appendix.

\section{Definitions and preliminaries}

The operator of the second-quantized relativistic spin-0 field in a static (ultrastatic) background is presented as
\begin{equation}\label{1}
    \hat{\Psi}(\textbf{x}, t) = \sum\hspace{-1.35em}\int\limits_\lambda \frac{1}{\sqrt{2E_\lambda}} \left[e^{-{\rm i} E_\lambda t} \psi_\lambda(\textbf{x}) \hat{a}_\lambda + e^{{\rm i} E_\lambda t} \psi_\lambda^\ast(\textbf{x}) \hat{b}^\dag_\lambda,\right]
\end{equation}
where natural units $\hbar = c = 1$ are used, $\hat a^\dag_\lambda$ and $\hat a_\lambda$ ($\hat b^\dag_\lambda$ and $\hat b_\lambda$) are the spin-0 particle (antiparticle) creation and destruction operators satisfying commutation relations
\begin{equation}
    [\hat{a}_\lambda,\hat{a}^\dag_{\lambda^\prime}]_- = [\hat{b}_\lambda,\hat{b}^\dag_{\lambda^\prime}]_- = \langle\lambda|\lambda^\prime\rangle,
\end{equation}
and $\lambda$ is the set of parameters (quantum numbers) specifying the state; wave functions $\psi_\lambda(\textbf{x})$ form a complete set of solutions to the stationary Fock-Klein-Gordon equation,
\begin{equation}\label{3}
    \bigl[-\mbox{\boldmath $\nabla$}^2 + m^2 + \xi R(\textbf{x})\bigr] \psi_\lambda(\textbf{x}) = E^2_\lambda \psi_\lambda(\textbf{x}),
\end{equation}
\mbox{\boldmath $\nabla$} is the covariant derivative involving the bundle connection, $R(\textbf{x})$ is the curvature scalar, $E_\lambda>0$ is the energy of the state, and symbol $\sum\hspace{-1em}\int\limits_\lambda$\; in \eqref{1} denotes summation over discrete and integration (with a certain measure) over continuous values of $\lambda$. The current which is induced in the vacuum is defined through the vacuum expectation value of the anticommutator of the field operators:
\begin{multline}\label{4}
    \textbf{j}(\textbf{x}) \equiv \frac{1}{2{\rm i}} \langle{\rm vac}|\Bigl\{ \bigl[\hat{\Psi}^\dag(\textbf{x},t), \mbox{\boldmath $\nabla$} \hat{\Psi}(\textbf{x},t)\bigr]_+ - \bigl[\mbox{\boldmath $\nabla$} \hat{\Psi}^\dag(\textbf{x},t),\hat{\Psi}(\textbf{x},t) \bigr]_+\Bigr\}|{\rm vac}\rangle =\\
    = -{\rm i} \sum\hspace{-1.35em}\int\limits_\lambda (2E_\lambda)^{-1} \Bigl\{\psi^\ast_\lambda(\textbf{x}) \bigl[\mbox{\boldmath $\nabla$} \psi_\lambda(\textbf{x})\bigr] - \bigl[\mbox{\boldmath $\nabla$} \psi_\lambda(\textbf{x})\bigr]^\ast \psi_\lambda(\textbf{x}) \Bigr\},
\end{multline}
where the vacuum is conventionally defined by relation
\begin{equation}
    \hat{a}_\lambda|{\rm vac}\rangle = \hat{b}_\lambda|{\rm vac}\rangle = 0.
\end{equation}

A straight infinitely long cosmic string in its rest frame is characterized by two parameters:

gauge flux
\begin{equation}\label{6}
    \Phi = \int\limits_{\rm core}d\mbox{\boldmath $\sigma$}\cdot \mbox{\boldmath $\partial$}\times \textbf{V}(\textbf{x})
\end{equation}

and tension (or linear density of mass)
\begin{equation}\label{7}
    M = \frac{1}{16\pi\c{G}} \int\limits_{\rm core}d\sigma \, R(\textbf{x}).
\end{equation}
Here $\textbf{V}(\textbf{x})$ is the vector potential of the gauge field corresponding to the spontaneously broken gauge symmetry, $\c{G}$ is the gravitational constant, and the integration is over the transverse section of the string core. Without loss of generality, the string core is assumed to have the form of a tube of radius $r_0$. A tube in $3$-dimensional space can be obviously generalized to a $(d-2)$ tube in $d$-dimensional space by adding extra $d-3$ dimensions as longitudinal ones.
Space outside the string core is locally flat ($R=0$) but non-Euclidean, with squared length element
\begin{equation}
    ds^2 = dr^2 + (1-4\c{G} M)^2 r^2 d\varphi + d\textbf{z}^2_{d-2},
\end{equation}
where $r$ and $\varphi$ are the polar coordinates on a surface which is transverse to the string, and $\textbf{z}_{d-2}$ are the Cartesian coordinates in the longitudinal directions. Such a space can be denoted as a conical one, since its transverse section is isometric to the surface of a cone with the deficit angle equal to $8\pi\c{G}M$. In general, the values of the deficit angle are bounded from above by $2\pi$, whereas they are unbounded from below (surplus angle can be arbitrarily large):
\begin{equation}\label{9}
    -\infty<8\pi\c{G}M<2\pi.
\end{equation}
If we return to real cosmic strings in $3$-dimensional space, then the tension (and, consequently, the deficit angle) surely has to be non-negative; moreover, the observation of discontinuities in the cosmic microwave background radiation imposes the stringent upper bound $M\lesssim 10^{-7}\c{G}^{-1}$ (see \cite{Bat}). However, negative values of $M$ and corresponding saddle-like conical spaces, as well as positive values of $M$ up to $(4\c{G})^{-1}$, can be of some physical interest. Various nanoconical structures arise in a diverse set of condensed matter systems known as the Dirac materials, ranging from honeycomb crystalline allotropes (graphene \cite{Ge}, silicene and germanene \cite{Cah}, phosphorene \cite{Liu}) to high-temperature cuprate superconductors \cite{Tsu} and topological insulators \cite{Qi}; in particular, a saddle-like conical space effectively emerges due to a radial disgyration in the A phase of superfluid He3, see \cite{Vol}.

Matter is assumed to interact with a cosmic string in the minimal way, i.e. the covariant derivative takes form $\mbox{\boldmath $\nabla$} = \mbox{\boldmath $\partial$} - {\rm i}\tilde{e} \textbf{V} + \frac{{\rm i}}{2} \mbox{\boldmath $\omega$}$, where $\tilde{e}$ is the appropriate coupling constant and
$\mbox{\boldmath $\omega$}$ is the affine connection. In the gauge with the only one nonvanishing component,
\begin{equation}
    V_\varphi = \frac{\Phi}{2\pi},
\end{equation}
the Laplace-Beltrami operator takes form
\begin{equation}
    \mbox{\boldmath $\nabla$}^2 = r^{-1}\frac{\partial}{\partial r} r \frac{\partial}{\partial r} + (1-4\c{G}M)^{-2} r^{-2} \biggl(\frac{\partial}{\partial\varphi} - \frac{{\rm i}\tilde{e}\Phi}{2\pi}\biggr)^2 + \left(\frac{\partial}{\partial \textbf{z}_{d-2}}\right)^2,
\end{equation}
and we obtain the following expression for a general solution to \eqref{3}, describing a state belonging to the continuous spectrum:
\begin{multline}\label{12}
    \psi_{kn\textbf{p}}(\textbf{x}) = (2\pi)^{(1-d)/2} e^{{\rm i} \textbf{p}\cdot\textbf{z}_{d-2}} e^{{\rm i}n\varphi} (1-4\c{G}M)^{-1/2}\\
    \times\left[\sin(\mu_n) J_{|n-\tilde{e}\Phi/(2\pi)|/\left(1-4\c{G}M\right)} (kr) + \cos(\mu_n) Y_{|n-\tilde{e}\Phi/(2\pi)|/\left(1-4\c{G}M\right)}(kr) \right],
\end{multline}
where $0<k<\infty$, $-\infty<p_j<\infty$ ($j=\overline{1,d-2}$), $E = \sqrt{\textbf{p}^2 + k^2 + m^2}$, $n\in\mathbb{Z}$ ($\mathbb{Z}$ is the set of integer numbers), $J_\rho(u)$ and $Y_\rho(u)$ are the Bessel functions of order $\rho$ of the first and second kinds. Parameter $\mu_n$ has to be determined from the boundary condition at $r=r_0$, see the next section. In the case of a vanishing transverse size of the string, relation
\begin{equation}\label{13}
\mu_n\big|_{r_0=0} = \pi/2
\end{equation}
is assumed, then the wave function in this case is regular at $r=0$, obeying orthonormality condition
\begin{equation}
    \int d^d \textbf{x} \sqrt{g} \psi^\ast_{kn\textbf{p}}(\textbf{x})\bigg|_{r_0=0} \psi_{k^\prime n^\prime \textbf{p}^\prime}(\textbf{x})\bigg|_{r_0=0} = \frac{\delta(k-k^\prime)}{k} \delta_{n,n^\prime}\delta^{d-2}(\textbf{p} - \textbf{p}^\prime).
\end{equation}

Substituting \eqref{12} into \eqref{4}, we  obtain  the induced vacuum current due to the contribution of the continuous spectrum, and this part  possesses an angular component as the only nonvanishing one,
\begin{equation}\label{15}
    j_\varphi^{(CS)}(r) = \int d^{d-2} \textbf{p} \int\limits_0^\infty \frac{dk\,k}{\sqrt{m^2 + \textbf{p}^2 + k^2}} \sum_{n\in\mathbb{Z}} \left(n - \frac{\tilde{e}\Phi}{2\pi}\right) |\psi_{kn\textbf{p}}(\textbf{x)}|^2.
\end{equation}
It is obvious from \eqref{12} and \eqref{15} that $j_\varphi^{(CS)}(r)$ is a periodic function of the string flux, $\Phi$ \eqref{6}, with a period equal to $2\pi/\tilde{e}$, i.e. it depends on quantity
\begin{equation}\label{16}
    F = \frac{\tilde{e}\Phi}{2\pi} - \biggl[\!\biggl[\frac{\tilde{e}\Phi}{2\pi} \biggr]\!\biggr],\quad 0\leq F <1
\end{equation}
and not on $\left[\!\left[\frac{\tilde{e}\Phi}{2\pi} \right]\!\right]$, where $[[u]]$ denotes the integer part of quantity $u$. Such a periodicity certainly is a manifestation of the renowned Aharonov-Bohm effect \cite{Ehre,Aha}. Defining quantity
\begin{equation}\label{17}
    \nu = (1-4\c{G}M)^{-1}
\end{equation}
and inserting \eqref{12} into \eqref{15}, we rewrite the latter as
\begin{multline}\label{18}
    j_\varphi^{(CS)}(r) = (2\pi)^{1-d} \int d^{d-2} \textbf{p} \int\limits_0^\infty \frac{dk\,k}{\sqrt{m^2 +  \textbf{p}^2 + k^2}} \sum_{n\in\mathbb{Z}} \nu(n-F) \bigl[\sin^2(\mu_n) J^2_{\nu|n-F|}(kr) \\
    + \sin(2\mu_n) J_{\nu|n-F|} (kr) Y_{\nu|n-F|} (kr) + \cos^2(\mu_n) Y^2_{\nu|n-F|} (kr) \bigr].
\end{multline}

If the angular component of the induced vacuum current is the only nonvanishing one, then the magnetic field strength in the longitudinal directions, $B^{3\ldots d}_{\rm I} (r)$, is also induced in the vacuum, as a consequence of the Maxwell equation,
\begin{equation}\label{19}
    \frac{r}{\nu} \, \frac{\partial}{\partial r} B^{3\ldots d}_{\rm I}(r) = -e j_\varphi(r),
\end{equation}
where $e$ is the electromagnetic coupling constant that in general differs from $\tilde{e}$. The total flux of the induced vacuum magnetic field is then given by expression
\begin{equation}\label{20}
    \Phi_{\rm I} = \nu^{-1}\int\limits_0^{2\pi}d\varphi \int\limits_{r_0}^\infty dr\,r B^{3\ldots d}_{\rm I}(r).
\end{equation}

 {Induced vacuum current, as well as consequent magnetic field, in the background of a vortex-type defect attracted attention several decades ago in view of its anticipated relevance to the Aharonov-Bohm effect. In the pioneering study in \cite{Ser,Alf,Gor,Fle,Par,Si6,Si7,SiB1,SiB2}, a simplifying unphysical assumption of the vanishing transverse size ($r_0=0$) and tension ($M=0$) of the vortex was employed. It was realized that the total induced vacuum magnetic flux is finite in the $d=2$ case and infinite in the physically interesting $d=3$ case. The problem of removing the above restriction, i.e. going over to $r_0>0$ and $M \neq 0$, was completely solved for the vacuum of a spin-1/2 quantized matter field in the $d=2$ case \cite{SiG} and in the $d=3$ case \cite{Si21}. In the present study, we consider the same problem for the vacuum of a spin-0 quantized matter field in the $d \geq 2$ case.}

\section{Self-adjointness of the Laplace-Beltrami operator and choice of boundary conditions}

Our attention is drawn to the Laplace-Beltrami operator, because, in the case of relativistic bosonic matter, the relevant quantum-mechanical operator is that of one-particle energy squared, see \eqref{3}. Defining a scalar product as
\begin{equation}
    (\tilde{\chi},\chi) = \int\limits_X d^d\textbf{x}\sqrt{g}\tilde{\chi}^\ast \chi,
\end{equation}
we get, using integration by parts,
\begin{equation}
    (\tilde{\chi},\mbox{\boldmath $\nabla$}^2 \chi) = (\mbox{\boldmath $\nabla$}^2 \tilde{\chi}, \chi) + \int\limits_{\partial X} d\mbox{\boldmath $\sigma$} \cdot [\tilde{\chi}^\ast (\mbox{\boldmath $\nabla$}\chi) - (\mbox{\boldmath $\nabla$} \tilde{\chi})^\ast \chi],
\end{equation}
where $\partial X$ is a hypersurface bounding the $d$-dimensional spatial region $X$. Operator $\mbox{\boldmath $\nabla$}^2$ is Hermitian (or symmetric  in mathematical parlance),
\begin{equation}
    (\tilde{\chi},\mbox{\boldmath $\nabla$}^2\chi) = (\mbox{\boldmath $\nabla$}^2\tilde{\chi},\chi),
\end{equation}
if relation
\begin{equation}\label{24}
    \int\limits_{\partial X} d\mbox{\boldmath $\sigma$}\cdot [\tilde{\chi}^\ast (\mbox{\boldmath $\nabla$}\chi) - (\mbox{\boldmath $\nabla$}\tilde{\chi})^\ast \chi] = 0
\end{equation}
holds. The latter can be satisfied in various ways by imposing different boundary conditions for $\chi$ and $\tilde{\chi}$. However, among the whole variety, there may exist a possibility that a boundary condition for $\tilde{\chi}$ is the same as that for $\chi$; then operator $\mbox{\boldmath $\nabla$}^2$ is self-adjoint. The action of a self-adjoint operator on functions of its domain of definition results in functions of the same domain only, and, therefore, a multiple action and functions of such an operator can be consistently defined. The spectral theorem (see, e.g., \cite{Ree1}) is valid for self-adjoint operators only, and this allows one to construct unitary exponents of them, see also \cite{Ree2}.

In the case of a connected boundary, condition \eqref{24} implies
\begin{equation}\label{25}
    \textbf{n}\cdot[\tilde{\chi}^\ast(\mbox{\boldmath $\nabla$}\chi) - (\mbox{\boldmath $\nabla$} \tilde{\chi})^\ast \chi]\big|_{\textbf{x} \in \partial X} = 0,
\end{equation}
where $\textbf{n}$ is the unit normal to boundary $\partial X$.
Defining
\begin{equation}
    \chi_{\pm} = \textbf{n}\cdot\mbox{\boldmath $\nabla$}\chi \pm \frac{\rm i}{\textbf{n}\cdot\textbf{x}}\chi,\qquad \tilde{\chi}_{\pm} = \textbf{n}\cdot\mbox{\boldmath $\nabla$}\tilde{\chi} \pm \frac{\rm i}{\textbf{n}\cdot\textbf{x}}\tilde{\chi},
\end{equation}
we rewrite \eqref{25} as
\begin{equation}
    \frac{\rm i}{2} \textbf{n}\cdot\textbf{x}(\tilde{\chi}^\ast_+ \chi_+ - \tilde{\chi}^\ast_- \chi_-)\big|_{\textbf{x} \in \partial X} = 0.
\end{equation}
The latter condition is satisfied by imposing linear conditions
\begin{equation}\label{28}
    (\chi_- - \Xi\chi_+)\big|_{\textbf{x} \in \partial X} = 0,\quad (\tilde\chi_- - \Xi\tilde\chi_+)\big|_{\textbf{x} \in \partial X} = 0,\quad |\Xi|^2 = 1.
\end{equation}
Using parametrization $\Xi = e^{2{\rm i}\theta}$, we rewrite \eqref{28} as
\footnote{Note that, in the case of a disconnected two-component boundary, the most general boundary condition depends on four parameters, see \cite {SiY}.}
\begin{equation}\label{29}
    \left[\left(\frac{\cos\theta}{\textbf{n}\cdot\textbf{x}} + \sin\theta\, \textbf{n}\cdot\mbox{\boldmath $\nabla$}\right)\chi\right]\bigg|_{\textbf{x} \in \partial X} = 0,\qquad \left[\left(\frac{\cos\theta}{\textbf{n}\cdot\textbf{x}} + \sin\theta\, \textbf{n}\cdot\mbox{\boldmath $\nabla$}\right)\tilde \chi\right]\bigg|_{\textbf{x} \in \partial X} = 0.
\end{equation}
One can recognize that \eqref{29} is actually the Robin boundary condition, with $\theta = 0$ corresponding to the Dirichlet condition (perfect reflectivity of the boundary) and
$\theta =\pm \pi/2$ corresponding to the Neumann condition (absolute rigidity of the boundary). The condition is periodic in the value of $\theta$ with the period equal to $\pi$, and the range of $\theta$ can be restricted to $-\pi/2\leq\theta<\pi/2$.

Parameter $\theta$ can be regarded as the self-adjoint extension parameter. It should be emphasized that the values of this parameter may in general vary from point to point on the boundary. In this respect the "number" of self-adjoint extension parameters is infinite, moreover, it is not countable but is of power of a continuum. This distinguishes the case of an extended boundary from the case of an excluded point, when the number of self-adjoint extension parameters is finite, being equal to $n^2$ for the deficiency index equal to $(n,n)$, see, e.g., \cite{Ree2}. Defining the quantum-mechanical current of matter as
\begin{equation}
    \textbf{J}_\lambda(\textbf{x}) = -{\rm i}\left\{\psi^\ast_\lambda(\textbf{x})\left[\mbox{\boldmath $\nabla$} \psi_\lambda(\textbf{x})\right] - \left[\mbox{\boldmath $\nabla$} \psi_\lambda(\textbf{x})\right]^\ast \psi_\lambda(\textbf{x}) \right\},
\end{equation}
we note that imposing the self-adjoint extension condition \eqref{29} on 
$\psi_\lambda$ results in the vanishing of the normal component of the current at the boundary,
\begin{equation}
    \textbf{n}\cdot\textbf{J}_\lambda(\textbf{x})\bigg|_{\textbf{x} \in \partial X} = 0.
\end{equation}
Thus the Robin boundary condition with the position-dependent boundary parameter is the most general one ensuring the impenetrability of a connected boundary for a spin-0 matter field.  {It should be noted that, in the case of a disconnected two-component boundary, the requirement of self-adjointness allows for a penetrable boundary, but the influx of quantum matter through one boundary component has to be equal to its outflux through another one, see \cite{SiY,Si}.}

In the case of a cosmic string, the boundary condition takes form \footnote{ {The impenetrability of the string core on the one hand is a consequence of the self-adjointness of the quantum-mechanical operator of energy squared. On the other hand, it is inevitable, since the vacuum of a spin-0 quantized matter field is not defined inside the string core.} }
\begin{equation}
    \left[\left(\cos\theta + \sin\theta\frac{r\partial}{\partial r}\right)\chi\right]\bigg|_{r=r_0} = 0,\qquad \left[\left(\cos\theta + \sin\theta\frac{r\partial}{\partial r}\right)\tilde\chi\right]\bigg|_{r=r_0} = 0.
\end{equation}
Imposing this condition on wave function \eqref{12}, we determine parameter $\mu_n$:
\begin{equation}\label{33}
    \tan\mu_n = -\frac{\left[\left(\cot\theta + r\frac{\partial}{\partial r}\right)Y_{\nu|n-F|} (kr) \right]\bigg|_{r=r_0}}{\left[\left(\cot\theta + r\frac{\partial}{\partial r}\right)J_{\nu|n-F|} (kr) \right]\bigg|_{r=r_0}}.
\end{equation}
As was already noted, the values of parameter $\theta$ in general depend on $\varphi$ and $\textbf{z}_{d-2}$. As a consequence, current \eqref{18} additionally depends on $\varphi
$ and $\textbf{z}_{d-2}$ via the dependence of $\mu_n$ ($n \in \mathbb{Z}$) on $\varphi$ and $\textbf{z}_{d-2}$. To be more precise, we assume the following ansatz for a solution to the stationary Fock-Klein-Gordon equation in the cosmic string background:
\begin{equation}
    \psi_{kn\textbf{p}}(\textbf{x}) = (2\pi)^{(1-d)/2} e^{i\textbf{p} \cdot \textbf{z}_{d-2}} e^{in\varphi} f_n(kr),
\end{equation}
where function $f_n(kr)$ is the solution to equation
\begin{equation}
    \left[ -\frac{1}{r}\frac{\partial}{\partial r} r \frac{\partial}{\partial r} + \frac{\nu^2}{r^2}(n-F)^2 - k^2 \right] f_n(kr) = 0
\end{equation}
and obeys boundary condition
\begin{equation}
    \left[\left(\cos\theta + \sin\theta \frac{r\partial}{\partial r}\right) f_n(kr)\right] \Big|_{r=r_0} = 0.
\end{equation}
Note also that, by taking the limit of $r_0 \rightarrow 0$ in \eqref{33}, we justify assumption \eqref{13}.

In the case of $\min\{F,1-F\}<\nu^{-1} \cot\theta<\infty$, in addition to solutions with $k>0$, see \eqref{12}, there are solutions with $k={\rm i}\kappa_n$, where $\kappa_n$ is determined by relation
\begin{equation}\label{37}
    \left[\left(\cos\theta + \sin\theta\frac{r\partial}{\partial r}\right)K_{\nu|n-F|} (\kappa_n r) \right]\Big|_{r=r_0}=0,
\end{equation}
$K_\rho(u)$ is the Macdonald function of order $\rho$. These solutions correspond to bound states on a surface which is transverse to a cosmic string,
\begin{multline}\label{38}
    \psi^{(BS)}_{\kappa_n n \textbf{p}} (\textbf{x}) = (2\pi)^{(1-d)/2} e^{{\rm i}\textbf{p}\cdot\textbf{z}_{d-2}} e^{{\rm i}n\varphi} \frac{\sqrt{2\nu}}{r_0} \\
    \times \left[K_{\nu|n-F|+1}(\kappa_n r_0) K_{\nu|n-F|-1}(\kappa_n r_0) - K^2_{\nu|n-F|}(\kappa_n r_0)\right]^{-1/2} K_{\nu|n-F|} (\kappa_n r).
\end{multline}
The dependence of $\kappa_n$ on the value of $\nu|n-F|$ for fixed values of $\theta$ is illustrated in Fig.\ref{fig:1}. For fixed values of $\theta$ and $\textbf{p}$, there is no more than one bound state for each value of $n$, and their energies, $E_n^{(BS)}=\sqrt{m^2+\textbf{p}^2 - \kappa_n^2}$, are in the gap below the continuum,
\begin{equation}
    \sqrt{m^2+\textbf{p}^2 - [\cot^2\theta - \nu^2(n-F)^2] r_0^{-2}} < E_n^{(BS)} < \sqrt{m^2+\textbf{p}^2}.
\end{equation}
Note that, in the limit of a vanishing transverse size of the string, wave function \eqref{38} vanishes at $\nu|n-F|>1$,  whereas it is novanishing at $\nu|n-F|<1$.\footnote{It should be noted that, in the case of an infinitely thin string, the requirement of self-adjointness of the Laplace-Beltrami operator allows for regular and irregular square-integrable at $r=0$ modes. The deficiency index can be ($n,n$) with $n=0,1,2,...$, depending on the values of $\nu$ and $F$. In particular, it is ($2,2$) for $\nu=1$, see, e.g., \cite{Ada,Dab}.}

A variation of $\theta$ with $\varphi$ can be moderate enough, so that a dependence of $\kappa_n$ on $\varphi$ can be neglected. Then, substituting \eqref{38} into \eqref{4}, we obtain 
\begin{multline}\label{40}
    j^{(BS)}_\varphi (r) = \frac{2(2\pi)^{1-d}}{r_0^2} \int d^{d-2} \textbf{p} \sum_{n\in\mathbb{Z}} \frac{\nu(n-F)}{\sqrt{m^2+\textbf{p}^2-\kappa_n^2}}\\
    \times \frac{\Theta(\cot\theta-\nu|n-F|) K^2_{\nu|n-F|}(\kappa_n r)}{K_{\nu|n-F|+1}(\kappa_n r_0) K_{\nu|n-F|-1} (\kappa_n r_0) - K^2_{\nu|n-F|}(\kappa_n r_0)},
\end{multline}
where
\begin{equation*}
    \Theta(u) = \left\{\begin{aligned}
    &1,&& u>0\\
    &0,&& u<0
    \end{aligned} \right\}
\end{equation*}
is the step function. Thus, the induced vacuum current in the cosmic string background is
\begin{equation}\label{41}
    j_\varphi(r) = j_\varphi^{(CS)} (r) + j_\varphi^{(BS)}(r),
\end{equation}
with $j_\varphi^{(CS)}$ and $j_\varphi^{(BS)}$ given by \eqref{18} and \eqref{40}, respectively.
It should be noted that the integral in \eqref{40}, as well as that in\eqref{18}, is divergent at $|\textbf{p}| \rightarrow \infty$ (logarithmically at $d=3$ and as a power at $d>3$). However, we shall show that these divergences cancel each other in the sum in \eqref{41}. Note also that a variation of $\theta$ with $\textbf{z}_{d-2}$ can be moderate enough, so that a violation of translational invariance along the longitudinal directions can be regarded as negligible. Then the induced vacuum magnetic field strength is in the longitudinal directions, as is given by \eqref{19}.

\section{Induced vacuum magnetic flux: analytic expressions}

We present induced vacuum current $j_\varphi(r)$ \eqref{41} as
\begin{equation}
    j_\varphi(r) = j_\varphi^{(a)}(r) + j_\varphi^{(b)}(r),
\end{equation}
where
\begin{equation}\label{43}
    j^{(a)}_\varphi (r) = (2\pi)^{1-d} \int d^{d-2}\textbf{p} \int\limits_0^\infty \frac{dk\,k}{\sqrt{m^2 + \textbf{p}^2+k^2}} \sum_{n\in\mathbb{Z}} \nu(n-F) J^2_{\nu|n-F|} (kr)
\end{equation}
and
\begin{multline}\label{44}
    j^{(b)}_\varphi(r) = (2\pi)^{1-d}\int d^{d-2}\textbf{p} \int\limits_0^\infty \frac{dk\,k}{\sqrt{m^2 + \textbf{p}^2+k^2}} \sum_{n\in\mathbb{Z}} \nu(n-F)\\
    \times \Bigl\{\cos^2(\mu_n) \bigl[Y_{\nu|n-F|}^2(kr) - J_{\nu|n-F|}^2(kr)\bigr] + \sin(2\mu_n) J_{\nu|n-F|}(kr) Y_{\nu|n-F|}(kr) \Bigr\} + j_\varphi^{(BS)}(r).
\end{multline}
The Bessel functions, $J_\rho(u)$ and $Y_\rho(u)$, in \eqref{43} and \eqref{44} are expressed
through the modified Bessel function, $I_\rho(u)$, and the Macdonald function, $K_\rho(u)$, and the integration is extended to negative values of $k$ as well, $-\infty<k<\infty$, see \cite{SiV6}. Such an integral can be regarded as the integral over the real axis in the complex $k$ plane. In the case of $j_\varphi^{(a)}(r)$ \eqref{43}, the integrand as a function of the complex $k$ variable possesses branching points at $k = \pm {\rm i}\sqrt{m^2+\textbf{p}^2}$, and the integration path over the real axis is continuously deformed to envelope a cut on the imaginary axis from ${\rm i}\sqrt{m^2+\textbf{p}^2}$ to ${\rm i}\infty$. In this way we get
\begin{multline}\label{45}
    j_\varphi^{(a)}(r) = \frac{4}{(2\pi)^d} \int d^{d-2}\textbf{p} \int\limits_{m_{|\textbf{p}|}}^\infty \frac{dq\,q}{\sqrt{q^2 - m_{|\textbf{p}|}^2}}  \sum_{n\in\mathbb{Z}} \nu(n-F) I_{\nu|n-F|}(qr)K_{\nu|n-F|}(qr),\\[-1em]
    m_{|\textbf{p}|} = \sqrt{m^2+\textbf{p}^2}.
\end{multline}
Using relation (see \cite{Prud})
\begin{equation*}
    I_\rho(u) K_\rho(u) = \frac12 \int\limits_0^\infty \frac{dy}y \exp\left(-\frac{u^2}{2y} - y\right) I_\rho(y)
\end{equation*}
and the Schl\"afli contour representation for $I_\rho(y)$, one can perform the summation over $n$ and then perform the integrations. The final result,
\begin{multline}\label{46}
    j_\varphi^{(a)}(r) = -\frac{8}{(4\pi)^{(d+1)/2}} \frac{m^{(d+1)/2}}{r^{(d-3)/2}} \biggl\{\frac{1}{2\pi}\int\limits_0^\infty du K_{(d+1)/2}[2mr\cosh(u/2)] \\
    \times \frac{\sinh(u/2)}{[\cosh(u/2)]^{(d-1)/2}} \frac{\sin(\nu F\pi)\sinh[\nu(1-F)u] - \sin[\nu(1-F)\pi]\sinh(\nu Fu)}{\cosh(\nu u) - \cos(\nu \pi)} \\
    + \frac1\nu \sum_{l=1}^{[\![\nu/2]\!]} K_{(d+1)/2}[2mr\sin(l\pi/\nu)] \frac{\cos(l\pi/\nu)}{[\sin(l\pi/\nu)]^{(d-1)/2}} \sin(2Fl\pi) \biggr\},
\end{multline}
was obtained in \cite{SiV9} for the case of $0<\nu\leqslant 2$, while additional terms appearing at $\nu>2$ were obtained in \cite{Bez1}. Further, we get
\begin{multline}\label{47}
    B^{3\ldots d(a)}_I (r) = e\nu \int\limits_r^\infty \frac{dr^\prime}{r^\prime} j^{(a)}_\varphi (r^\prime) = -\frac{4e\nu}{(4\pi)^{(d+1)/2}} \left(\frac{m}{r}\right)^{(d-1)/2}\\
    \times\Biggl\{ \frac{1}{2\pi} \int\limits_0^\infty du\, K_{(d-1)/2} [2mr\cosh (u/2)] \frac{\sinh(u/2)}{[\cosh(u/2)]^{(d+1)/2}}\\
    \times \frac{\sin(\nu F\pi) \sinh[\nu(1-F)u] - \sin[\nu(1-F)\pi] \sinh(\nu Fu)}{\cosh(\nu u) - \cos(\nu \pi)}\\
    + \frac{1}{\nu} \sum_{l=1}^{[\![\nu/2]\!]} K_{(d-1)/2} [2mr\sin(l\pi/\nu)] \frac{\cos(l\pi/\nu)}{[\sin(l\pi/\nu)]^{(d+1)/2}} \sin(2Fl\pi) \Biggr\}
\end{multline}
and
\begin{multline}\label{48}
    \Phi_I^{(a)} = \frac{2\pi}{\nu} \int\limits_{r_0}^{\infty} dr\,r\,B_I^{3\ldots d(a)}(r) = -\frac{e}{(4\pi)^{(d-1)/2}} \left(\frac m{r_0}\right)^{(d-3)/2} \\
    \times \Biggl\{ \frac{1}{2\pi} \int\limits_0^\infty du\, K_{(d-3)/2} [2mr_0 \cosh(u/2)] \frac{\sinh(u/2)}{[\cosh(u/2)]^{(d+3)/2}} \Biggr.\\
    \times \frac{\sin(\nu F\pi)\sinh[\nu(1-F)u] - \sin[\nu(1-F)\pi]\sinh(\nu Fu)}{\cosh(\nu u) - \cos(\nu \pi)}\\
    \Biggl.+ \frac{1}{\nu} \sum_{l=1}^{[\![\nu/2]\!]} K_{(d-3)/2} [2mr_0 \sin(l\pi/\nu)] \frac{\cos(l\pi/\nu)}{[\sin(l\pi/\nu)]^{(d+3)/2}} \sin(2Fl\pi)\Biggr\}.
\end{multline}
In the case of the vanishing string tension, 
$\nu=1$, the results for $j^{(a)}_\varphi(r)$ and $B_I^{3\ldots d(a)}(r)$ were obtained more than two decades ago, see \cite{SiB1, SiB2},
\begin{equation}
    j^{(a)}_\varphi (r)\Big|_{\nu=1} = \frac{32\sin(F\pi)}{(4\pi)^{(d+3)/2}} \frac{m^{(d+1)/2}}{r^{(d-3)/2}} \int\limits_1^\infty  \frac{dv}{v^{(d+1)/2}} K_{(d+1)/2} (2mrv) \sinh[(2F-1) {\rm arccosh}(v)]
\end{equation}
and
\begin{equation}
    B_I^{3\ldots d(a)} (r)\Big|_{\nu=1} = \frac{16e\sin(F\pi)}{(4\pi)^{(d+3)/2}} \left(\frac mr\right)^{(d-1)/2} \int\limits_1^\infty \frac{dv}{v^{(d+3)/2}} K_{(d-1)/2} (2mrv) \sinh[(2F-1){\rm arccosh}(v)],
\end{equation}
whereas the flux is given by expression
\begin{equation}\label{51}
    \Phi_I^{(a)}\Big|_{\nu=1} = \frac{4e\sin(F\pi)}{(4\pi)^{(d+1)/2}} \left(\frac{m}{r_0}\right)^{(d-3)/2} \int\limits_1^\infty \frac{dv}{v^{(d+5)/2}} K_{(d-3)/2} (2mr_0v) \sinh[(2F-1){\rm arccosh}(v)].
\end{equation}

For generic $\nu$, we find out that flux $\Phi_I^{(a)}$  \eqref{48} in the limit of a vanishing transverse size of the string is finite in the $d=2$ case only, while otherwise a divergence occurs:
\begin{align}
    &\lim\limits_{r_0\to0}\Phi_I^{(a)}|_{d=2} = \frac{e}{4m} I_3(F,\nu),\label{52}\\
    &\Phi_I^{(a)}|_{d=3} \underset{r_0\to0}{=} \frac{e}{4\pi}[-\ln(mr_0)] I_3(F,\nu),
\end{align}
and
\begin{equation}\label{54}
    \Phi_I^{(a)}|_{d>3} \underset{r_0\to0}{=} \frac{e\Gamma(\frac{d-3}{2})}{2(4\pi)^{(d-1)/2}} r_0^{3-d} I_d(F,\nu),
\end{equation}
where
\begin{multline}
    I_d(F,\nu) = -\frac{1}{2\pi} \int\limits_0^\infty du \frac{\sinh(u/2)}{[\cosh(u/2)]^d} \frac{\sin(\nu F\pi)\sinh[\nu(1-F)u] - \sin[\nu(1-F)\pi]\sinh(\nu Fu)}{\cosh(\nu u) - \cos(\nu\pi)}\\
    - \frac{1}{\nu} \sum_{l=1}^{[\![\nu/2]\!]} \frac{\cos(l\pi/\nu)}{[\sin(l\pi/\nu)]^d} \sin(2Fl\pi),\label{55}
\end{multline}
$\Gamma(u)$ is the Euler gamma function. The integration and summation in \eqref{55} can be performed in the case of the odd $d$ values, see Appendix. In particular, we obtain
\begin{align}
    &\lim\limits_{r_0\to0}\Phi^{(a)}_I|_{d=2} = \frac{e}{6m} \left(F-\frac12\right)F(1-F)\nu^2,\label{56}\\
    &\Phi^{(a)}_I|_{d=3} \underset{r_0\to0}{=} \frac{e}{6\pi}[-\ln(mr_0)]\left(F-\frac12\right)F(1-F)\nu^2,\label{57}\\
    &\Phi^{(a)}_I|_{d=5} \underset{r_0\to0}{=} \frac{e}{12(2\pi)^2}r_0^{-2}\left(F-\frac12\right)F(1-F)\nu^2\biggl\{\frac13+\frac15\Bigl[\frac13+F(1-F)\Bigr]\nu^2\biggr\},\label{58}\\
    &\Phi^{(a)}_I|_{d=7} \underset{r_0\to0}{=} \frac{e}{180(2\pi)^3} r_0^{-4}\left(F-\frac12\right)F(1-F)\nu^2\biggl\{\frac43 + \Bigl[\frac13 + F(1-F)\Bigr]\nu^2 \nonumber\\
    &\qquad\qquad+ \frac17\Bigl[\frac13 + F(1-F) + F^2(1-F)^2\Bigr]\nu^4\biggr\},\\
    &\Phi^{(a)}_I|_{d=9} \underset{r_0\to0}{=} \frac{e}{810(2\pi)^4} r_0^{-6} \left(F-\frac12\right)F(1-F)\nu^2 \biggl\{4+\frac{49}{15}\Bigl[\frac13+F(1-F)\Bigr]\nu^2 \nonumber\\
    &\qquad\qquad + \frac23\Bigl[\frac13 + F(1-F) + F^2(1-F)^2\Bigr]\nu^4 + \frac1{15}\Bigl[\frac13 + F(1-F)\nonumber\\
    &\qquad\qquad+\frac{10}9 F^2(1-F)^2 + \frac59 F^3(1-F)^3\Bigr]\nu^6\biggr\},\label{60}
\end{align}
and so on; note the positive definiteness of all terms in curly brackets in \eqref{58}--\eqref{60}. Note also that \eqref{55} can be calculated exactly for all $d$ values in the case of a vanishing string tension, see \cite{Gor16},
\begin{equation}
    I_d(F,1) = \frac{\sin(F\pi)}{2\sqrt\pi}\left(F-\frac12\right)\frac{\Gamma(\frac{d-1}{2}+F)\Gamma(\frac{d+1}{2}-F)}{\Gamma(\frac d2 + 1) \Gamma(\frac{d+1}{2})}.
\end{equation}

Turning now to the $\theta$-dependent piece of the induced vacuum current, see $j_\varphi^{(b)}(r)$ \eqref{44}, the integral over $k$ is transformed into the integral over a contour enveloping the upper imaginary semiaxis on the complex $k$ plane. In this way we get
\begin{multline}\label{62}
    j_\varphi^{(b)}(r) = \frac{8}{(2\pi)^{d+1}} \int d^{d-2}\textbf{p} \sum_{n\in\mathbb{Z}} \nu(n-F) \sum_{\pm}\biggl\{\int\limits_0^{m_{|\textbf{p}|}} \frac{dq\,q}{\sqrt{m_{|\textbf{p}|}^2 - q^2}} e^{\mp {\rm i}\nu|n-F|\pi} \Bigl[\cos^2(\mu_n)\Big|_{k=\pm {\rm i}q} \\
    \pm \frac {\rm i}2 \sin(2\mu_n)\Big|_{k=\pm {\rm i}q} \Bigr] K_{\nu|n-F|}^2(qr) + \int\limits_{m_{|\textbf{p}|}}^\infty \frac{dq\,q}{\sqrt{q^2 - m_{|\textbf{p}|}^2}} e^{\mp {\rm i}\nu|n-F|\pi} \Bigl[\pm {\rm i}\cos^2(\mu_n)\Big|_{k=\pm {\rm i}q} \\
    - \frac12 \sin(2\mu_n)\Big|_{k=\pm {\rm i}q} \Bigr] K^2_{\nu|n-F|}(qr) \biggr\} + j_\varphi^{(BS)}(r), \quad m_{|\textbf{p}|} = \sqrt{m^2 + \textbf{p}^2}.
\end{multline}
In view of relation
\begin{multline}\label{63}
    \sum_{\pm} e^{\mp {\rm i}\nu|n-F|\pi}\Bigl[\cos^2(\mu_n)\Big|_{k=\pm {\rm i}q} \pm \frac {\rm i}2 \sin(2\mu_n)\Big|_{k=\pm {\rm i}q} \Bigr] = -\frac{\pi^2}{\kappa_n r_0^2}\\
    \times \frac{\Theta(\cot\theta - \nu|n-F|) \delta(q-\kappa_n)}{K_{\nu|n-F|+1}(\kappa_n r_0) K_{\nu|n-F|-1}(\kappa_n r_0) - K^2_{\nu|n-F|}(\kappa_n r_0)},
\end{multline}
the contribution of the integral over $q$ from $0$ to $m_{|\textbf{p}|}$ cancels $j_\varphi^{(BS)}(r)$ \eqref{40}. Using relation
\begin{equation}
    \sum_{\pm} e^{\mp {\rm i}\nu|n-F|\pi}\Bigl[\pm {\rm i}\cos^2(\mu_n)\Big|_{k=\pm {\rm i}q} - \frac 12 \sin(2\mu_n)\Big|_{k=\pm {\rm i}q} \Bigr] = -\pi C_{\nu|n-F|}(\theta,qr_0),
\end{equation}
where
\begin{equation}\label{65}
    C_\rho(\theta,v) = \frac{[(\cot\theta + v\partial_v)I_\rho(v)]}{[(\cot\theta + v\partial_v)K_\rho(v)]},
\end{equation}
we obtain the following expression for the remaining piece, after integration over $\textbf{p}$,
\begin{equation}\label{66}
    j_\varphi^{(b)}(r)=-\frac{8}{(4\pi)^{(d+1)/2}}\frac{1}{\Gamma(\frac{d-1}{2})} \int\limits_m^\infty dq\,q (q^2-m^2)^{(d-3)/2} \sum_{n\in\mathbb{Z}} \nu(n-F) C_{\nu|n-F|}(\theta,qr_0) K^2_{\nu|n-F|}(qr);
\end{equation}
note that a zero in the denominator of $C_\rho(\theta,v)$ \eqref{65} is to be treated with the principal value prescription. It should be noted also that terms with $\nu|n-F| < 1$ in the integrand of $j_\varphi^{(BS)}(r)$ \eqref{40}, unlike terms with $\nu|n-F| > 1$, are finite in the limit of a vanishing transverse size of the string. However, as has been just remarked, they, as well as others, are canceled out, and the
$\theta$-dependent piece of the current, as is clear from  \eqref{65} and \eqref{66}, is vanishing in this limit,
\begin{equation}\label{67}
\lim\limits_{r_0\to0} j_\varphi^{(b)} (r) = 0.
\end{equation}

Further, we get the $\theta$-dependent piece of the induced vacuum magnetic field strength,
\begin{multline}
    B_I^{3\ldots d(b)}(r) = e\nu \int\limits_{r}^{\infty} \frac{dr^\prime}{r^\prime} j_\varphi^{(b)}(r^\prime) = -\frac{4e\nu}{(4\pi)^{(d+1)/2}} \frac{1}{\Gamma(\frac{d-1}{2})} \\
    \times \int\limits_m^\infty dq\,q(q^2-m^2)^{(d-3)/2} \sum_{n\in\mathbb{Z}} {\rm sgn} (n-F) C_{\nu|n-F|}(\theta,qr_0) \bigl[K^2_{\nu|n-F|}(qr) + qr W_{\nu|n-F|}(qr) \bigr],
\end{multline}
where ${\rm sgn}(u)=\Theta(u)-\Theta(-u)$ is the sign function and
\begin{equation}
    W_\rho(v) = K_\rho (v) \frac{d}{d\rho} K_{\rho-1} (v) - K_{\rho-1} (v) \frac{d}{d\rho} K_\rho (v);
\end{equation}
this piece also vanishes in the
$r_0  \rightarrow  0$ limit.
Thus, the $\theta$-independent pieces,  $j^{(a)}_\varphi (r)$ \eqref{46}
and  $B_I^{3\ldots d(a)}(r)$ \eqref{47}, can be regarded as corresponding to the case of an infinitely thin cosmic string.

As to the appropriate flux,
\begin{equation}
    \Phi_I^{(b)} = \frac{1}{\nu} \int\limits_0^{2\pi} d\varphi \int\limits_{r_0}^\infty dr\,r\,B_I^{3\ldots d(b)} (r),
\end{equation}
we obtain the following expression for it:
\begin{multline}\label{71}
    \Phi_I^{(b)} = -\frac{2e}{(4\pi)^{(d+1)/2}} \frac{r_0^{3-d}}{\Gamma(\frac{d-1}{2})} \int\limits_0^{2\pi} d\varphi \int\limits_{mr_0}^{\infty} dv\,v(v^2-m^2r_0^2)^{(d-3)/2} \sum_{n\in\mathbb{Z}} {\rm sgn}(n-F)
     C_{\nu|n-F|}(\theta,v)\\ \times \left[\nu|n-F| K_{\nu|n-F|+1} (v) K_{\nu|n-F|-1} (v) - (\nu|n-F| + 1) K_{\nu|n-F|}^2 (v) - v W_{\nu|n-F|} (v) \right].
\end{multline}
The total induced vacuum magnetic flux,
\begin{equation}\label{72}
\Phi_I=\Phi^{(a)}_I + \Phi^{(b)}_I,
\end{equation}
decreases as the transverse size of the string increases, $r_0  \rightarrow \infty$. Since, as a consequence of \eqref{67}, $\Phi^{(b)}_I$ vanishes in the limit of the vanishing transverse size of the string, the behavior of the total flux in this limit is governed by that of $\Phi^{(a)}_I$, and relations \eqref{52}--\eqref{54} and, consequently, \eqref{56}--\eqref{60} are relevant for
$\Phi_I$. In the case of
$\nu^{-1}\cot\theta<\min\{F,1-F\}$ (in particular, in the cases of Dirichlet and Neumann boundary conditions), the total flux as a function of $r_0$ does not change sign, being finite for $d=2$ and for $d>2$, except $r_0=0$. In the case of $\nu^{-1}\cot\theta > |n'-F|$ ($n'\in\mathbb{Z}$), the total flux can be rewritten as
\begin{multline}\label{73}
    \Phi_I = -\frac{2e}{(4\pi)^{(d+1)/2}} \frac{r_0^{3-d}}{\Gamma(\frac{d-1}{2})} \int\limits_0^{2\pi} d\varphi \int\limits_{mr_0}^{\infty} dv\,v(v^2-m^2r_0^2)^{(d-3)/2} \\ \times \sum_{n'\in\mathbb{Z}} {\rm sgn}(n'-F)\Theta(\cot\theta-\nu|n'-F|)
     C_{\nu|n'-F|}(\theta,v)\\ \times \left[\nu|n'-F| K_{\nu|n'-F|+1} (v) K_{\nu|n'-F|-1} (v) - (\nu|n'-F| + 1) K_{\nu|n'-F|}^2 (v) - v W_{\nu|n'-F|} (v) \right]
    + \tilde{\Phi}_I,
\end{multline}
where terms with the principal value prescription are explicitly exhibited as a finite sum over $n'$. Coefficient $C_{\nu|n'-F|}(\theta,v)$ changes sign in the vicinity of point $v_{n'}$:
\begin{equation}\label{74}
    C_{\nu|n'-F|}(\theta,v)\Big|_{v \sim  v_{n'}}  = - \left(v - v_{n'}\right)^{-1}\left\{v_{n'}\left[K_{\nu|n'-F|+1}(v_{n'}) K_{\nu|n'-F|-1}(v_{n'}) - K^2_{\nu|n'-F|}(v_{n'})\right]\right\}^{-1},
\end{equation}
where $v_{n'}$ is a root of equation, cf. \eqref{37},
\begin{equation}\label{75}
    \left[\left(\cos\theta + \sin\theta\frac{v\partial}{\partial v}\right)K_{\nu|n'-F|} (v) \right]\Big|_{v=v_{n'}}=0.
\end{equation}
As a consequence of the sign change of $C_{\nu|n'-F|}(\theta,v)$, the total flux in the case of $2 \leq d \leq 3$, as a function of $r_0$, becomes infinite at points $r_0=m^{-1}v_{n'}$. These peculiarities will be studied in more detail in the next section. Note that the peaks at points $r_0=m^{-1}v_{n'}$ are absent in the case of $d > 3$, and thus, in this case, the behavior of terms with a zero in the denominator in \eqref{71} is qualitatively the same as that of otherones.

\section{Induced vacuum magnetic flux: numerical results}

As is clear from \eqref{48} and \eqref{71}, the induced vacuum magnetic flux is odd under change $F \rightarrow 1-F$, where $F$ is defined by \eqref{6}. In the $d=2$ case, neglecting the transverse size of a cosmic string, one obtains \eqref{56} for $\Phi_I^{(a)}$ and zero for $\Phi_I^{(b)}$; thus, the maximal absolute value for the flux is attained at two points which are symmetric with respect to $F=1/2$,
\begin{equation}\label{76}
F_{\pm} = \frac12 \left(1 \pm \frac{1}{\sqrt{3}}\right). 
\end{equation}
Such a behavior of the flux as a function of $F$ remains qualitatively the same, when the transverse size of the string is taken into account, although the position of the maximum in the flux absolute value is shifted, depending on $r_0$, $\nu$, and $\theta$. To illustrate this, we display the $F$ dependence of the dimensionless flux in the $d=2$ and $d=3$ cases at $r_0=10^{-2}m^{-1}$ and several values of $\nu$ and 
$\theta$ in Fig.\ref{cfig:Fdependence}.  

Fixing $F=F_+$, let us now consider the behavior of induced vacuum magnetic flux as a function of the transverse size of the string. Results at $\cot\theta \leq 0$ in the case of $d=2$ and several values of $\nu$ are presented in Fig.\ref{fig:2}. As boundary conditions vary from Dirichlet ($\theta=0$) to Neumann ($\theta=-\pi/2$), the flux monotonically increases in the region of small $mr_0$; note a bend near $mr_0=0$ in the case of $\theta=-\pi/2$. The region where flux is somewhat essential increases with the increase of $\nu$. A situation is qualitatively the same in the case of $d=3$, see Fig.\ref{fig:3}, with a distinction that flux diverges at $mr_0 \rightarrow 0$, see \eqref{57}. In the realistic case of $\nu \approx 1$, the values of $e^{-1}\Phi_I$ exceeding $10^{-3}$ are reached at
$mr_0 < 5 \cdot 10^{-2}$ for the Dirichlet condition and at $mr_0< 25 \cdot 10^{-2}$ for the Neumann one; note that the case of $\nu = 1$ was considered for the Dirichlet condition in \cite{Gor16} and for the Neumann one in \cite{Gor22}.

The situation drastically changes at $0 < \cot\theta < \infty$. The results for flux in the case of $d=2$ at several values of $\theta$ and $\nu$ are presented in Figs.\ref{fig:5}-\ref{fig:6}. As positive $\theta$ decreases starting from the value of $\pi/2$, more and more terms with a zero in the denominator contribute, see \eqref{73}. Before integration, as well as after it, these terms change sign, see \eqref{74}, at the points of zeros which are determined by \eqref{75}. The integration starts from $mr_0$, and each time, as increasing $mr_0$ passes point $v_{n'}$, the principal value prescription becomes decompensated just to the right of these points, giving rise to infinite peaks in the integral. As $\nu$ decreases, the number of peaks increases: more values of $n'$ satisfy inequality $\nu^{-1}\cot\theta > |n'-F|$. The situation looks somewhat different in the case of $d=3$, and the appropriate results for flux are presented in Figs.\ref{fig:8}-\ref{fig:9}. The behavior of integrands containing factors with a zero in the denominator is qualitatively the same with the only distinction that an overall factor of $(v^2-m^2r_0^2)^{-1/2}$ is missing in this case. Therefore, a contribution of interval $mr_0< v < v_{n'}$ might not be sufficient to outweigh a contribution of the opposite sign from $v > v_{n'}$. As a consequence, the flux can be of the same sign from both sides of the infinite peaks at $mr_0 = v_{n'}$. 

To illustrate this, let us consider the case of $\theta=3\pi/8$, $F=F_+$, and $\nu=1$, when the sum over $n'$ in \eqref{73} reduces to the only one term, $n'=1$,  
\begin{equation}\label{77}
G_{d}(mr_0) = e m^{d-3}\int\limits_{mr_0}^{\infty} dv\,g_d(v),
\end{equation}
where
\begin{multline}\label{78}
	g_d(v) = -\frac{v[v^2(mr_0)^{-2}-1]^{(d-3)/2}}{(4\pi)^{(d-1)/2}\Gamma(\frac{d-1}{2})} \, \, C_{F_-}\left(\frac{3\pi}{8},v \right)   \\
	\times \left[F_- K_{1+F_-} (v) K_{F_+} (v) - (1+F_-) K_{F_-}^2 (v) - v W_{F_-} (v) \right].
\end{multline}
We display both \eqref{78} and \eqref{77} at $d=2$ and at $d=3$ on Fig.\ref{fig:FirstTerm}. Although the behavior of $g_2(v)$ and $g_3(v)$ in the vicinity of $v_1$ is qualitatively the same, behavior of integrals, $G_{2}(mr_0)$ and $G_{3}(mr_0)$, differs due to an integrable divergence of $g_2(v)$ at $mr_0$. Note a symmetry with respect to point $v=v_1$ in the $d=3$ case, see Fig.\ref{fig:FirstTerm} (b) and Fig.\ref{fig:FirstTerm} (d). This signifies that relation $(v-v_1) g_3(v) = 0.0152$ holds with high precision for interval $0.07 <  v < 0.09$. Note also that $g_d(v)$ at $d>3$ is evidently integrable at $v=v_1=mr_0$.

Returning to the total induced vacuum magnetic flux in the $d=3$ case, see Figs.\ref{fig:3},\ref{fig:8}-\ref{fig:9}, we note that the infinite peak of positive sign to the right of point $mr_0=0$ exists, although it is hardly visible on some plots due to its narrowness; the numerical calculations start from the minimal value of $mr_0$ equal to $10^{-3}$, and the width of the peak can be less than this value in some cases.

As to the $d \geq 4$ case, the numerical analysis shows that integrals corresponding to separate terms in the sum over $n$ in \eqref{71} are infinite due to divergence at $v \rightarrow \infty$, see, in particular, \eqref{77} at $d \geq 4$. However, due to the sign changing, these divergencies are cancelled upon summation of the whole series.

\section{Summary}

In the present paper, we have considered a magnetic field that is induced by a cosmic string in the ground state of quantum relativistic bosonic (scalar or pseudoscalar) matter. The transverse size of the string is taken into account, and the string is obviously generalized to a ($d-2$) tube in $d$-dimensional space by adding extra $d-3$ dimensions as longitudinal ones. The most general boundary conditions ensuring the impenetrability of matter to the interior of the string is shown to be the Robin condition with one parameter, $\theta$, varying arbitrarily from point to point of the boundary. Provided that the variation of $\theta$ is moderate enough, we find that a current circulating around the string and a magnetic field strength directede along the string are induced in the ground state; they decrease exponentially at large distances from the string. We also find the total induced ground state magnetic flux, which is given by \eqref{72} with $\Phi_I^{(a)}$ and $\Phi_I^{(b)}$ given by \eqref{48} and \eqref{71}, respectively. These results provide a field-theoretical realization of the Aharonov-Bohm effect \cite{Ehre,Aha}, since they depend on $F$ \eqref{16}, i.e., they depend periodically on gauge flux $\Phi$ \eqref{6} with period equal to $2\pi/\tilde{e}$. The ground state characteristics are smooth continuous functions of $F$, vanishing at $F=0, \frac12, 1$ and being odd under change $F \rightarrow 1-F$. They depend on parameter $\nu$ \eqref{17}, which is related to string tension $M$ \eqref{7}; the range of $\nu$, in general, is $0 < \nu < \infty$, see \eqref{9}.

It should be noted that, for $0 < \cot\theta < \infty$, in addition to the continuum state solutions to the Fock-Klein-Gordon equation in the cosmic string background, there are solutions corresponding to  bound states on a surface that is transverse to the string. We show in the process of calculation of vacuum characteristics that their direct contribution is canceled, see \eqref{62} and \eqref{63}. However, the bound states reveal themselves in the manner of a smoking gun, through zeros in the denominators of coefficients $C_\rho(\theta,v)$ \eqref{65}, which are to be treated with the principal value prescription when integrated over $v$. The appearance of these zeros results in a thoroughgoing distinction of the case of $0 < \cot\theta < \infty$ from that of $\cot\theta \leq 0$.

The numerical analysis of the expression for induced vacuum magnetic flux, obtained in Section 4, is carried out in Section 5. Restricting ourselves to constant values of $\theta$, we find that the flux decreases significantly at large values of $mr_0$. In the case of $\cot\theta \leq 0$, the flux is of the same sign, as $mr_0$ varies from $0$ to $\infty$, see Figs.\ref{fig:2} and \ref{fig:3}; it is finite at $d=2$ and diverges logarithmically at $d=3$ in the limit of $mr_0 \rightarrow 0$. In contrast to this, the flux in the case of $0 < \cot\theta < \infty$ is a sign changing function of $mr_0$, see Figs.\ref{fig:5}--\ref{fig:9}. Moreover, it diverges at points $mr_0=v_{n'}$, where $v_{n'}$ are the above-mentioned zeros in the denominators, see \eqref{74} and \eqref{75}; the number of these points is determined by inequality $\nu^{-1}\cot\theta > |n'-F|$. The position of the rightmost infinite peak of the flux is shifted to region $mr_0 > 1$, as $\nu^{-1}\cot\theta$ increases; for instance, this happens at $\frac{\pi}{4} > \theta > \frac{\pi}{8}$ in the case of $F=F_{\pm}$ [where $F_{\pm}$ is given by \eqref{76}] and $\frac12 < \nu < 2$.

Note that, as is clear from the numerical results, the minimal absolute values for the induced vacuum magnetic flux in the cosmic string background are attained with the use of the Dirichlet boundary condition. Hereof, appealing in a somewhat sense to an analogue of the Occam's razor principle, we can eliminate the variety of boundary conditions by giving preference to that of Dirichlet, as the most plausible one in view of its minimal effect on the vacuum. However, more stringent arguments are given below that allow us to restrict the set of boundary conditions with more definiteness.    

The transverse size of a cosmic string is of the order of correlation length, $r_0 \sim m_H^{-1}$, where $m_H$ is the energy scale of spontaneous symmetry breaking, i.e., the mass of the appropriate Higgs boson. It looks hardly plausible that a topological defect (cosmic string) influences surrounding quantum matter with the matter particle mass, $m$, exceeding the Higgs particle mass, $m_H$. For instance, a cosmic string that is
formed at the grand unification scale can polarize the vacuum of quantum matter in the electroweak model, whereas a would-be cosmic string corresponding to the electroweak symmetry breaking has no impact on the vacuum of quantum matter in the grand unified model. The more implausible is an enormous influence, i.e., infinite peaks in the induced vacuum magnetic flux at $m > m_H$. This reasoning allows us to exclude the range of boundary parameters corresponding to $0 < \cot\theta < \infty$ as that leading to unphysical consequences.

Thus, we are left with the range of boundary parameters corresponding to $\cot\theta \leq 0$, when the induced vacuum magnetic flux attains visible values at $m << m_H$, see Figs.\ref{fig:2} and \ref{fig:3}. The effect is minimal for the case of the Dirichlet condition, $\theta=0$, slightly and monotonically increasing, as the boundary parameter gradually evolves in value to the Neumann condition, $\theta=-\pi/2$. This result has to be compared with that for the case of magnetic field that is induced by a cosmic string in the vacuum of quantum relativistic fermionic matter, see \cite{SiG,Si21}. In the latter case, the most general boundary condition involves one at $d=2$ \cite{SiG} and four at $d=3$ \cite{Si21} parameters varying arbitrarily from point to point of the boundary, and the requirement of finiteness for the induced vacuum magnetic flux removes completely the ambiguity in the choice of boundary conditions. We conclude that the impact of a topological defect (cosmic string) on quantum matter differs significantly for bosons and fermions. If bosons and fermions are assigned to a representation of a supersymmetry group, then this supersymmetry is violated by the vacuum effects in the cosmic string background.

\section*{Acknowledgments}

The work of Yu.A.S. was supported by the National Academy of Sciences of Ukraine (Project No. 0122U000886) and by the Special Research Program URF 2022 of the Erwin Schr\"{o}dinger International Institute for Mathematics and Physics. He thanks E.Pallante for fruitful discussion and critical remarks.

\newpage

\setcounter{equation}{0} \setcounter{figure}{0}
\renewcommand{\theequation}{A.\arabic{equation}}
\section*{Appendix. Case of a thin cosmic string}

\newcounter{p}

\begin{figure}[h!]
    \centering\includegraphics[width=0.7\textwidth]{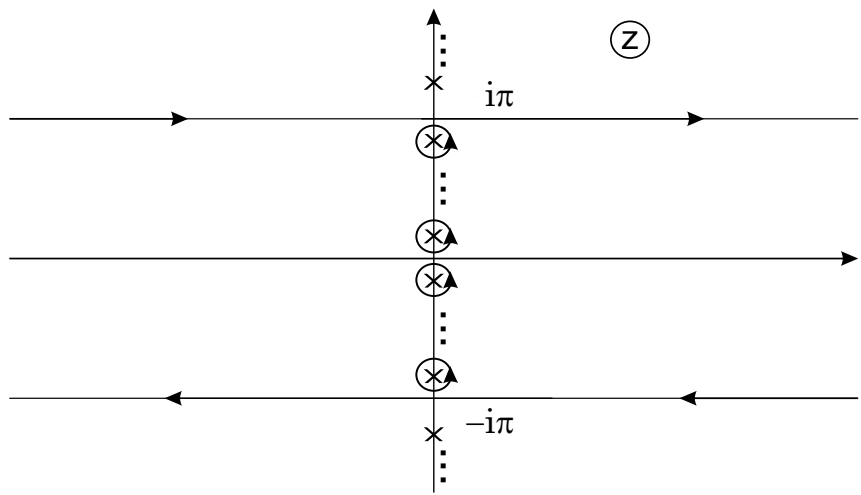}
    \caption{Singularities of the integrand in \eqref{A1} on the complex $z$ plane out of the origin are simple poles on the imaginary axis; they are denoted by crosses. Contour $C$ consists of two horizontal lines and circles around the simple poles with $0<|{\rm Im}\,z|<\pi$.}
    \label{fig:A1}
\end{figure}

\indent Only the pieces  with superscript $(a)$ are relevant for this case. Starting from $j^{(a)}_\varphi(r)$ \eqref{45}, we recall that summation over $n$ is performed by using the Schl\"afli contour integral representation for $I_\rho(u)$, see \cite{SiV9} for details. Generically, we obtain
\begin{multline}\label{A1}
    j^{(a)}_\varphi(r) = \frac{1}{(4\pi)^{(d+1)/2}} \frac{m^{(d+1)/2}}{r^{(d-3)/2}} \frac{1}{2\pi{\rm i}} \int\limits_C \frac{dz}{\Bigl[\sqrt{-\sinh^2(z/2)}\,\Bigr]^{(d+1)/2}} K_{(d+1)/2} [2mr\sqrt{-\sinh^2(z/2)}]\\
    \times \frac{\sinh(z) \sinh\bigl[\nu\bigl(F-\frac12\bigr)z\bigr]}{\sinh(\nu z/2)},
\end{multline}
where contour $C$ results from merging two different Schl\"afli contours for the modified Bessel functions of orders $\nu(n-F)>0$ and $-\nu(n-F)>0$, see Fig.\ref{fig:A1}; the contribution of horizontal lines yields the integral over $u$ in \eqref{46}, while the contribution of circles yields the sum over $l$ in \eqref{46}. Further, we get
\begin{multline}
    B_I^{3\ldots d(a)}(r) = \frac{e\nu}{2(4\pi)^{(d+1)/2}} \left(\frac{m}{r}\right)^{(d-1)/2} \frac{1}{2\pi{\rm i}} \int\limits_C \frac{dz}{\Bigl[\sqrt{-\sinh^2(z/2)}\,\Bigr]^{(d+3)/2}}\\
    \times K_{(d-1)/2}\Bigl[2mr\sqrt{-\sinh^2(z/2)}\Bigr] \frac{\sinh(z) \sinh\bigl[\nu\bigl(F-\frac12\bigr)z\bigr]}{\sinh(\nu z/2)}
\end{multline}
and
\begin{multline}\label{A3}
    \Phi_I^{(a)} = \frac{e}{8(4\pi)^{(d-1)/2}} \left(\frac{m}{r_0}\right)^{(d-3)/2} \frac{1}{2\pi{\rm i}} \int\limits_C \frac{dz}{\Bigl[\sqrt{-\sinh^2(z/2)}\,\Bigr]^{(d+5)/2}}\\
    \times K_{(d-3)/2}\Bigl[2mr_0\sqrt{-\sinh^2(z/2)}\Bigr] \frac{\sinh(z) \sinh\bigl[\nu\bigl(F-\frac12\bigr)z\bigr]}{\sinh(\nu z/2)}.
\end{multline}
Taking the asymptotics of $\Phi_I^{(a)}$ \eqref{A3} at $r_0\to0$, we obtain \eqref{52}--\eqref{54}, where $I_d(F,\nu)$ is presented as
\begin{equation}
    I_d(F,\nu) = \frac{1}{16\pi{\rm i}} \int\limits_C \frac{dz}{\Bigl[\sqrt{-\sinh^2(z/2)}\,\Bigr]^{d+1}} \frac{\sinh(z) \sinh\bigl[\nu\bigl(F-\frac12\bigr)z\bigr]}{\sinh(\nu z/2)}.
\end{equation}
In the case of odd values of $d$, a singularity of the integrand at $z=0$ is an isolated pole. Contour $C$ in this case can be continuously deformed to encircle the origin, and we get
\begin{equation}
    I_{2n+1}(F,\nu) = \frac{(-1)^n}{8\pi{\rm i}} \ointctrclockwise dz\frac{\cosh(z/2) \sinh\bigl[\nu\bigl(F-\frac12\bigr)z\bigr]}{\Bigl[\sinh(z/2)\Bigr]^{2n+1} \sinh(\nu z/2)}.
\end{equation}
Taking a residue of the simple pole at the origin, we obtain \eqref{56}--\eqref{60}.

\newpage

\begin{figure}[h!]  \centering\includegraphics[width=0.5\textwidth]{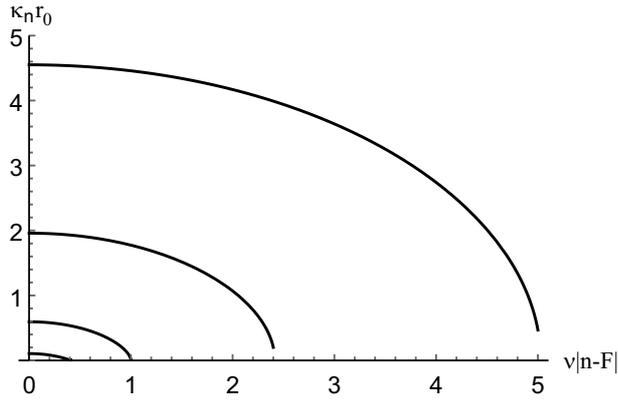}
    \caption{The value of the root of equation \eqref{37} as a function of $\nu|n-F|$: curves from top to bottom correspond to $\theta=\pi/16,\ \pi/8,\ \pi/4,\ 3\pi/8$.}
    \label{fig:1}
\end{figure}
\begin{figure}[h!]
    \centering
    \includegraphics[width=\textwidth]{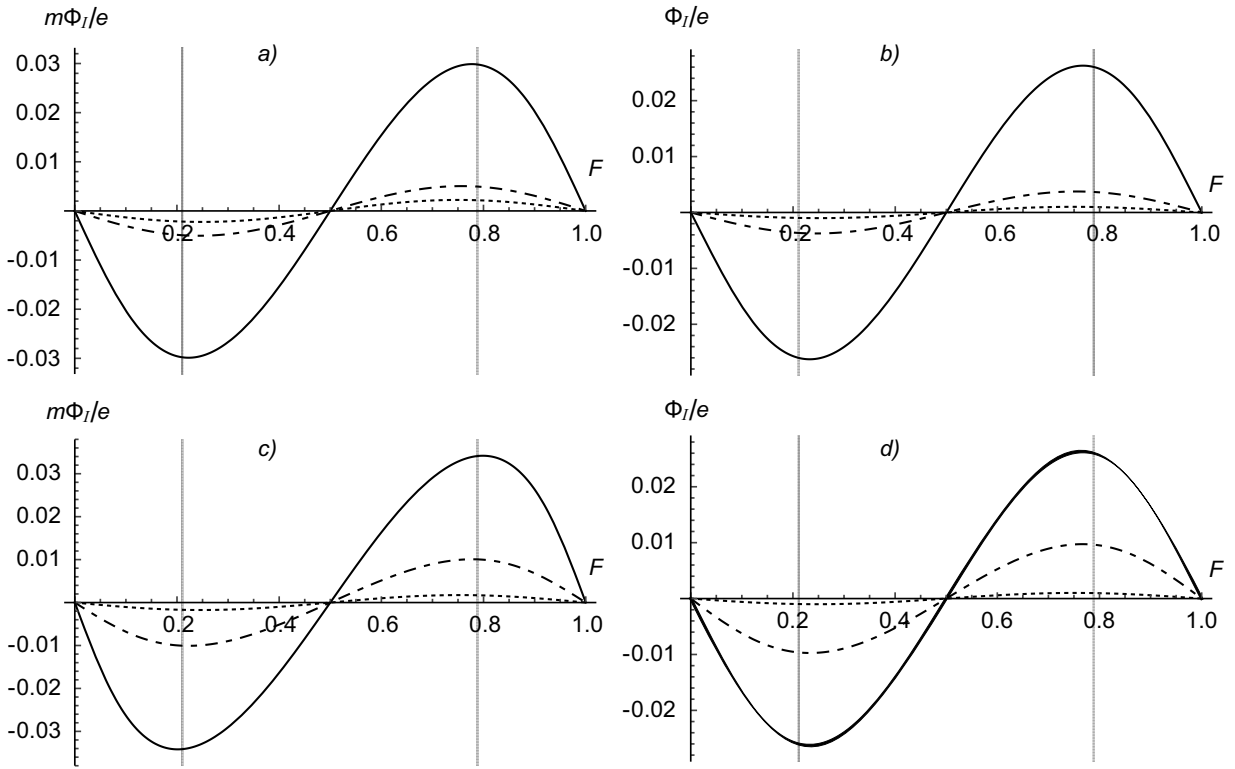}
    \caption{Dimensionless induced vacuum magnetic flux at $m r_0=10^{-2}$ as a function of $F$:
    a) $\theta=-\pi/4$, $d=2$, b) $\theta=-\pi/4$, $d=3$, c) $\theta=-\pi/2$, $d=2$, d) $\theta=-\pi/2$, $d=3$. Parameter $\nu$ takes values $2$, $1$, and $1/2$ for solid,  dashed-dotted, and dashed lines, correspondingly; the values for the case of $\nu=1/2$ in a) and b) are  multiplied by 10; vertical lines correspond to $F=F_{\pm}$.}
    \label{cfig:Fdependence}
\end{figure}

\begin{figure}[h!]
    \centering
    \includegraphics[width=0.9\textwidth]{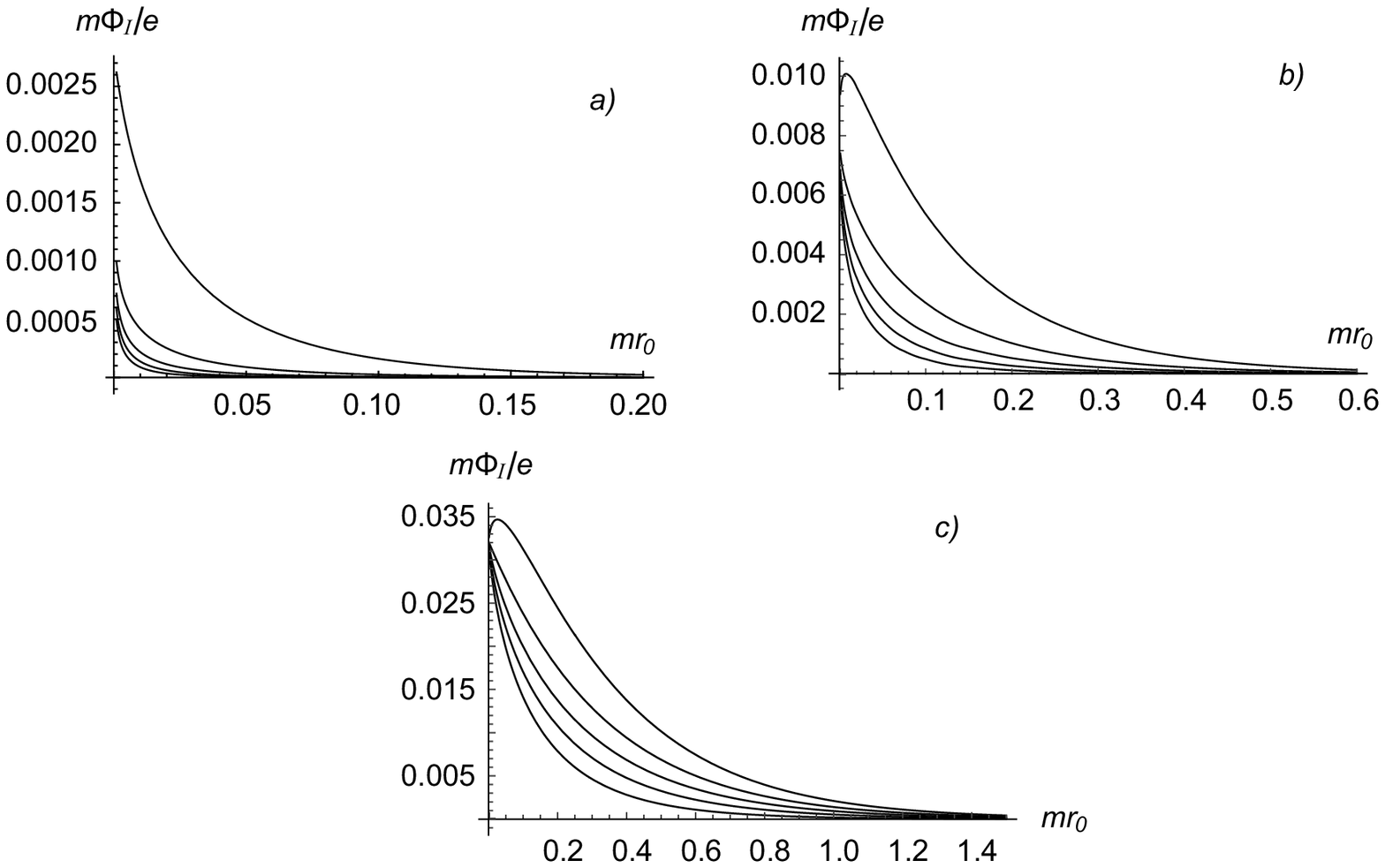}
    \caption{Dimensionless vacuum flux at $\cot\theta \leq 0$, $F=F_+$, and $d=2$: a) $\nu=1/2$, b) $\nu=1$, c) $\nu=2$. Parameter $\theta$ takes values $-\pi/2$, $-3\pi/8$, $-\pi/4$, $-\pi/8$, $0$ from top to bottom.}
    \label{fig:2}
\end{figure}

\begin{figure}[h!]
    \centering
    \includegraphics[width=0.9\textwidth]{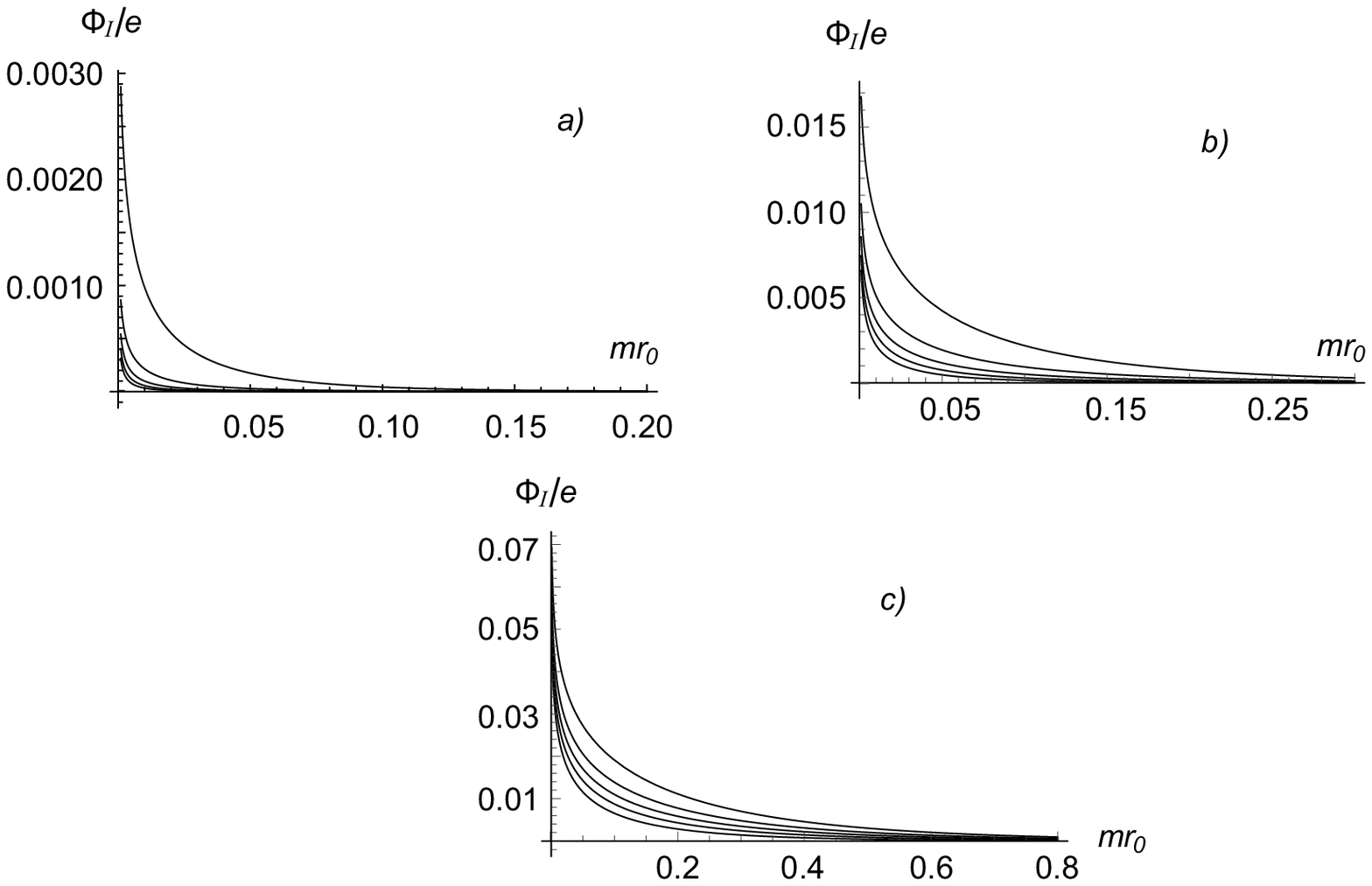}
    \caption{Dimensionless vacuum flux at $\cot\theta \leq 0$, $F=F_+$, and $d=3$: a) $\nu=1/2$, b) $\nu=1$, c) $\nu=2$. Parameter $\theta$ takes values $-\pi/2$, $-3\pi/8$, $-\pi/4$, $-\pi/8$, $0$ from top to bottom.}
    \label{fig:3}
\end{figure}

\begin{figure}[h!]
    \centering
    \includegraphics[width=\textwidth]{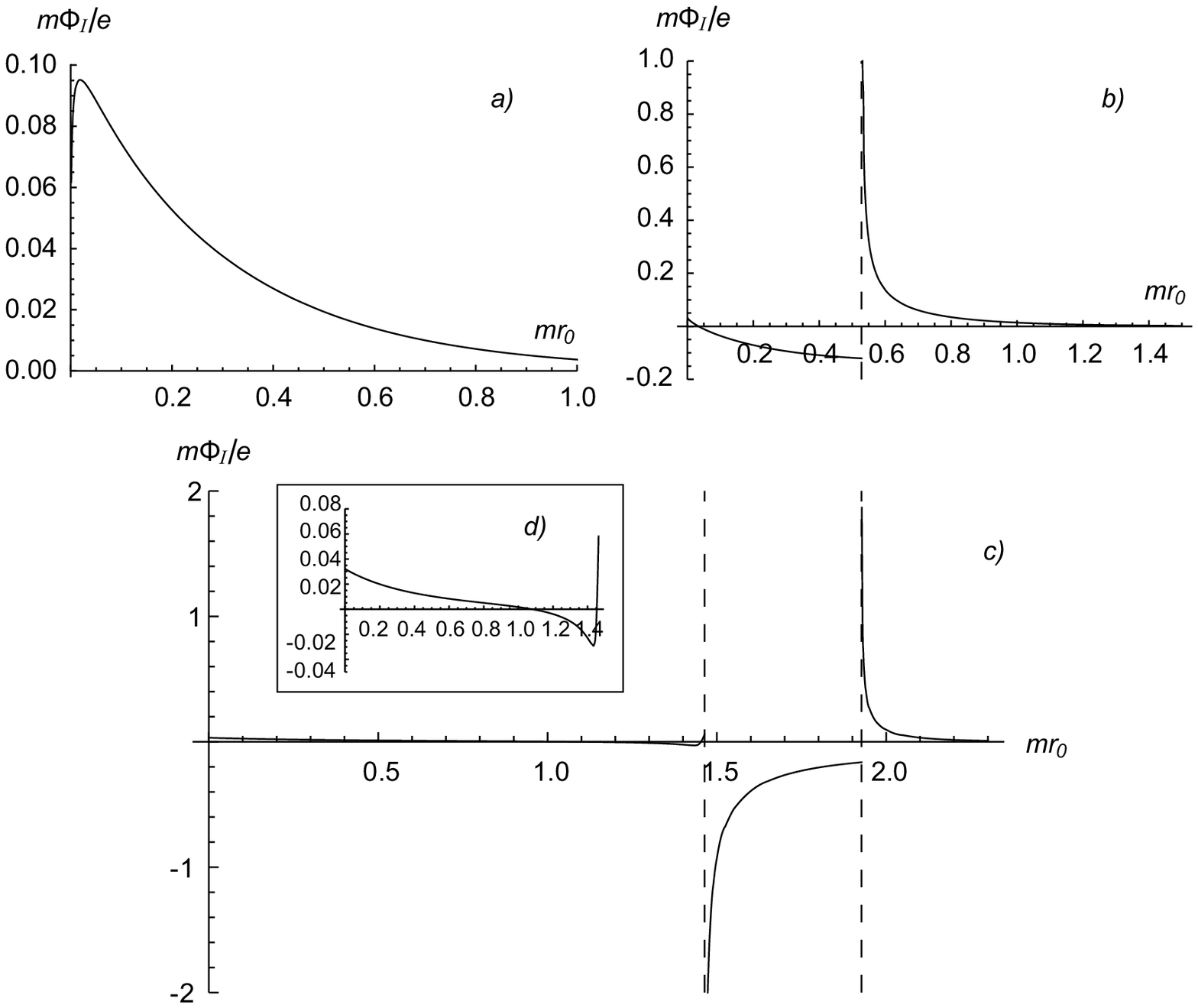}
    \caption{Dimensionless vacuum flux at $0 < \cot\theta < \infty$, $F=F_+$, $d=2$, and $\nu=2$: a) $\theta=3\pi/8$, b) $\theta=\pi/4$, c) $\theta=\pi/8$, d) is a zoomed enlargedpart of c).}
    \label{fig:5}
\end{figure}
\begin{figure}[h!]
    \centering
    \includegraphics[width=\textwidth]{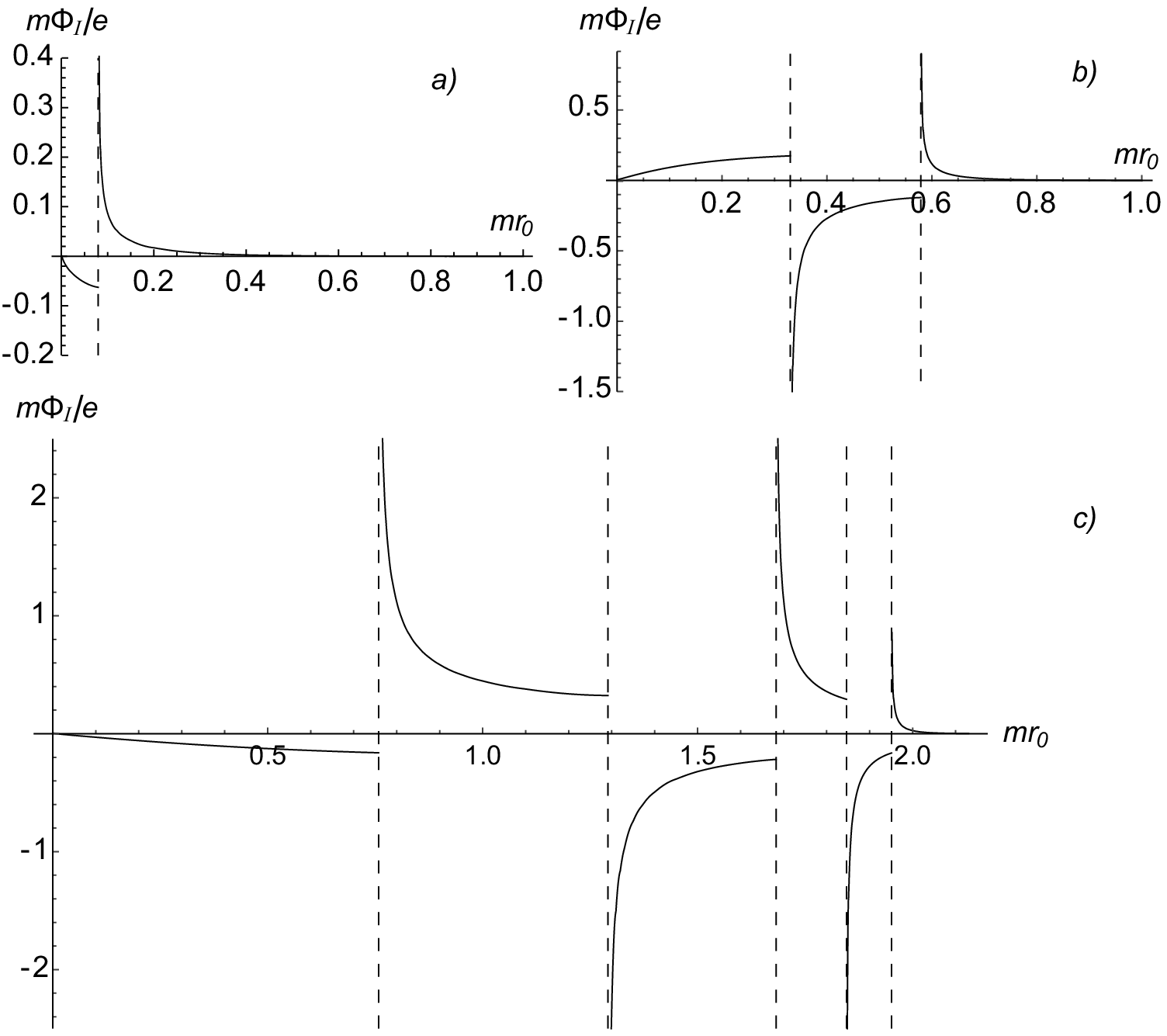}
    \caption{Dimensionless vacuum flux at $0 < \cot\theta < \infty$, $F=F_+$, $d=2$, and $\nu=1$: a) $\theta=3\pi/8$, b) $\theta=\pi/4$, c) $\theta=\pi/8$.}
    \label{fig:4}
\end{figure}

\begin{figure}[h!]
    \centering
    \includegraphics[width=\textwidth]{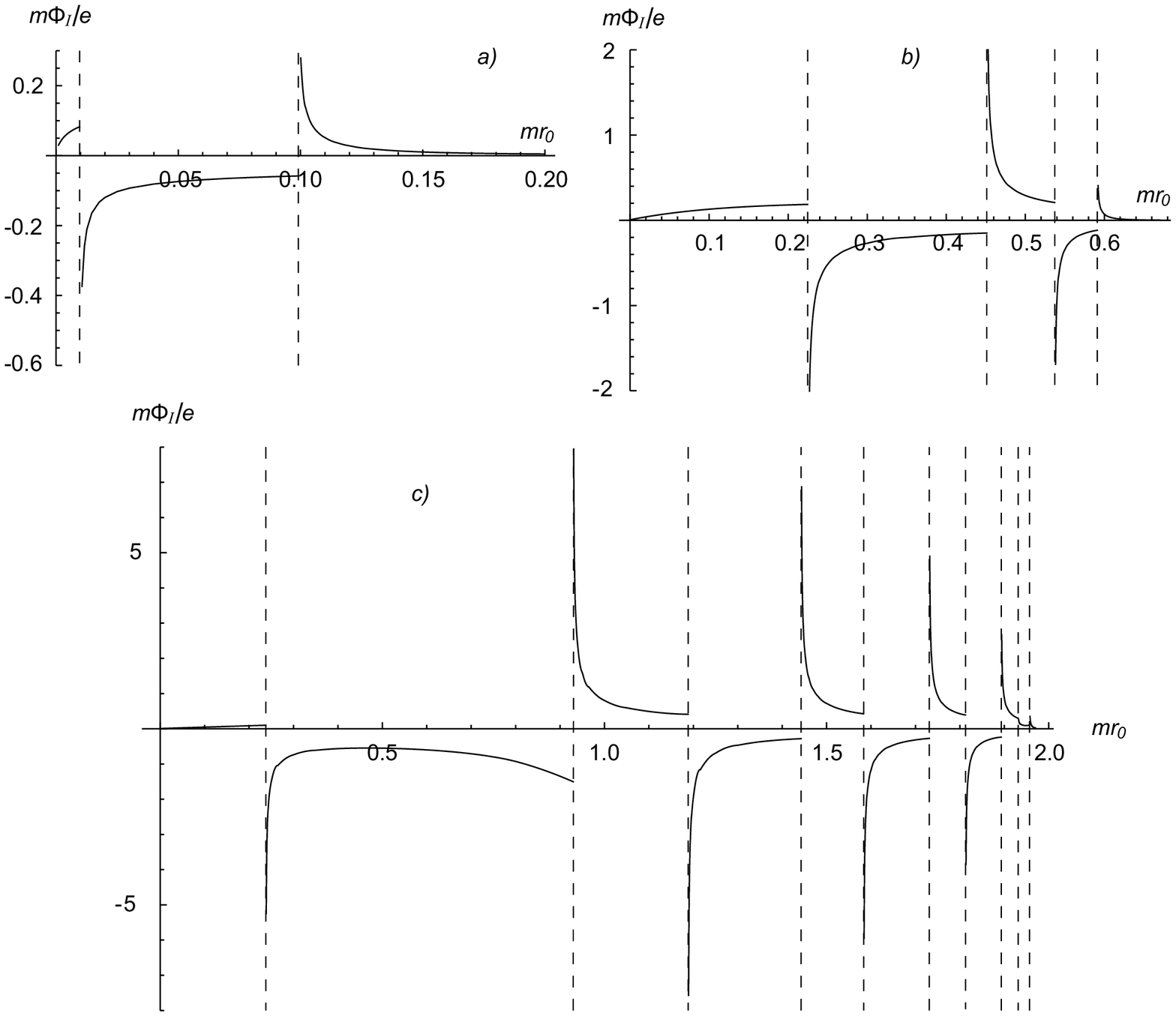}
    \caption{Dimensionless vacuum flux at $0 < \cot\theta < \infty$, $F=F_+$, $d=2$, and $\nu=1/2$: a) $\theta=3\pi/8$, b) $\theta=\pi/4$, c) $\theta=\pi/8$.}
    \label{fig:6}
\end{figure}

\begin{figure}[h!]
    \centering
    \includegraphics[width=0.9\textwidth]{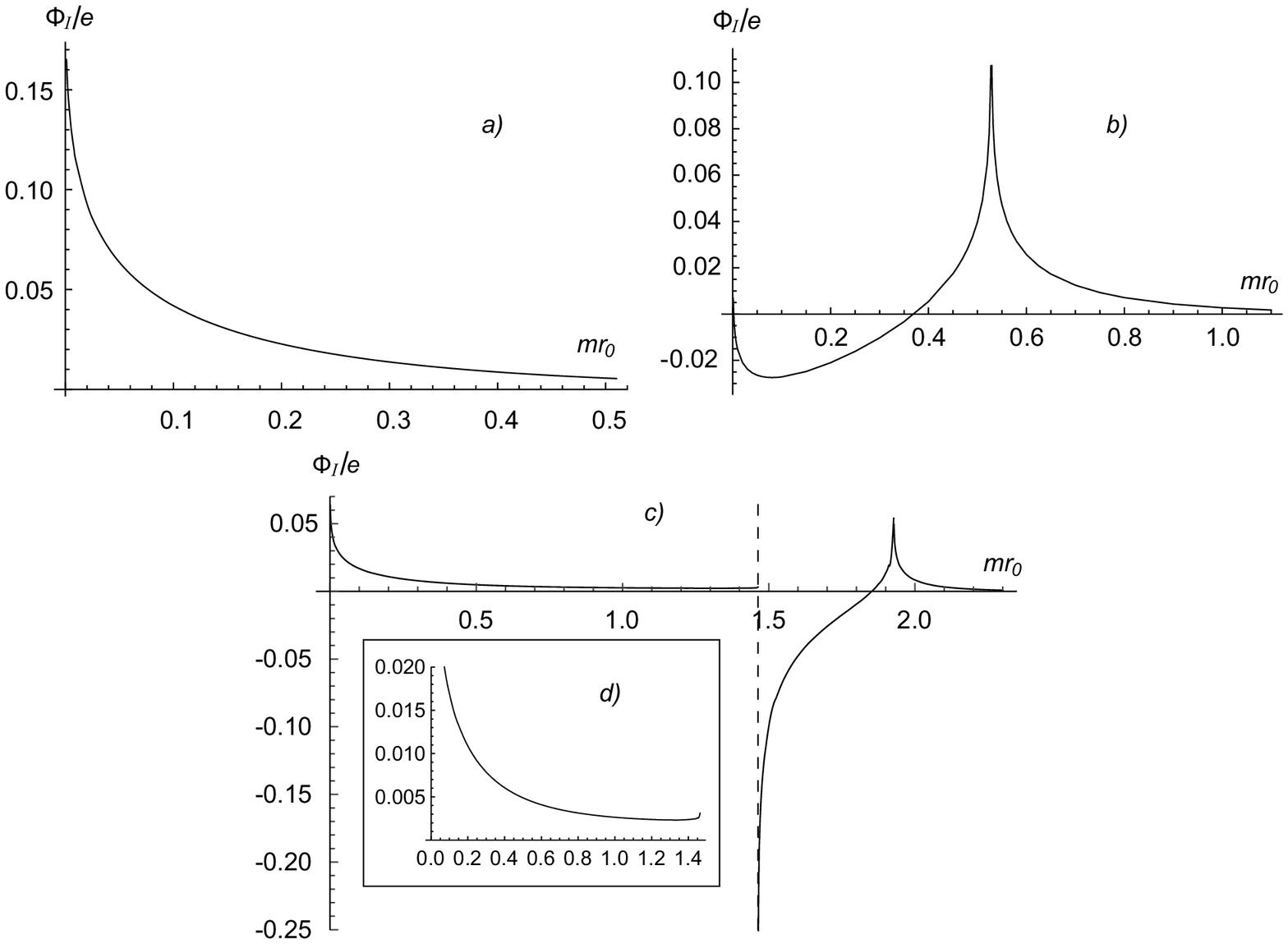}
    \caption{Dimensionless vacuum flux at $0 < \cot\theta < \infty$, $F=F_+$, $d=3$, and $\nu=2$: a) $\theta=3\pi/8$, b) $\theta=\pi/4$, c) $\theta=\pi/8$, d) is a zoomed enlarged part of c).}
    \label{fig:8}
\end{figure}
\begin{figure}[h!]
    \centering
    \includegraphics[width=0.9\textwidth]{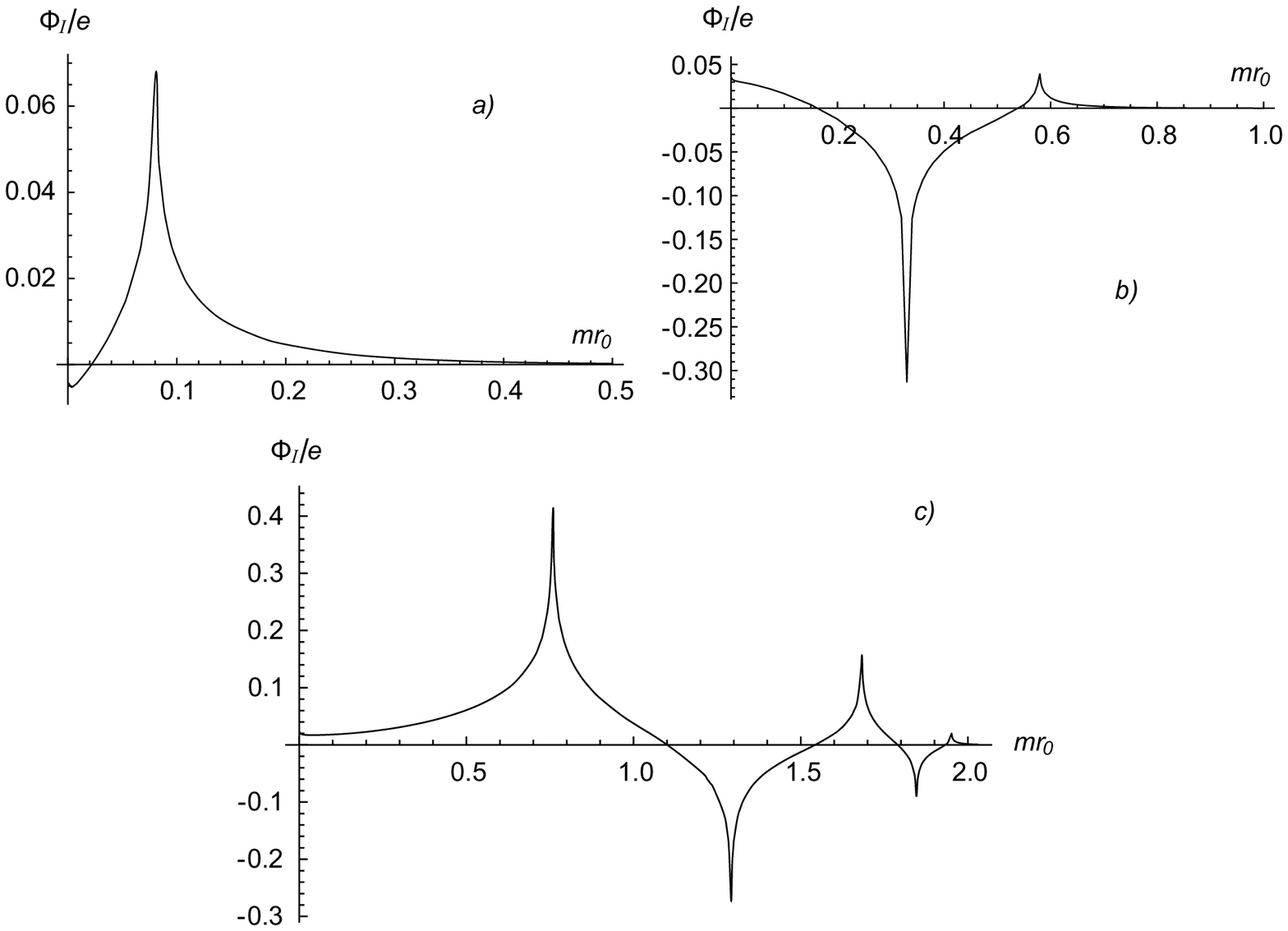}
    \caption{Dimensionless vacuum flux at $0 < \cot\theta < \infty$, $F=F_+$, $d=3$, and $\nu=1$: a) $\theta=3\pi/8$, b) $\theta=\pi/4$, c) $\theta=\pi/8$.}
    \label{fig:7}
\end{figure}

\begin{figure}[h!]
    \centering
    \includegraphics[width=0.9\textwidth]{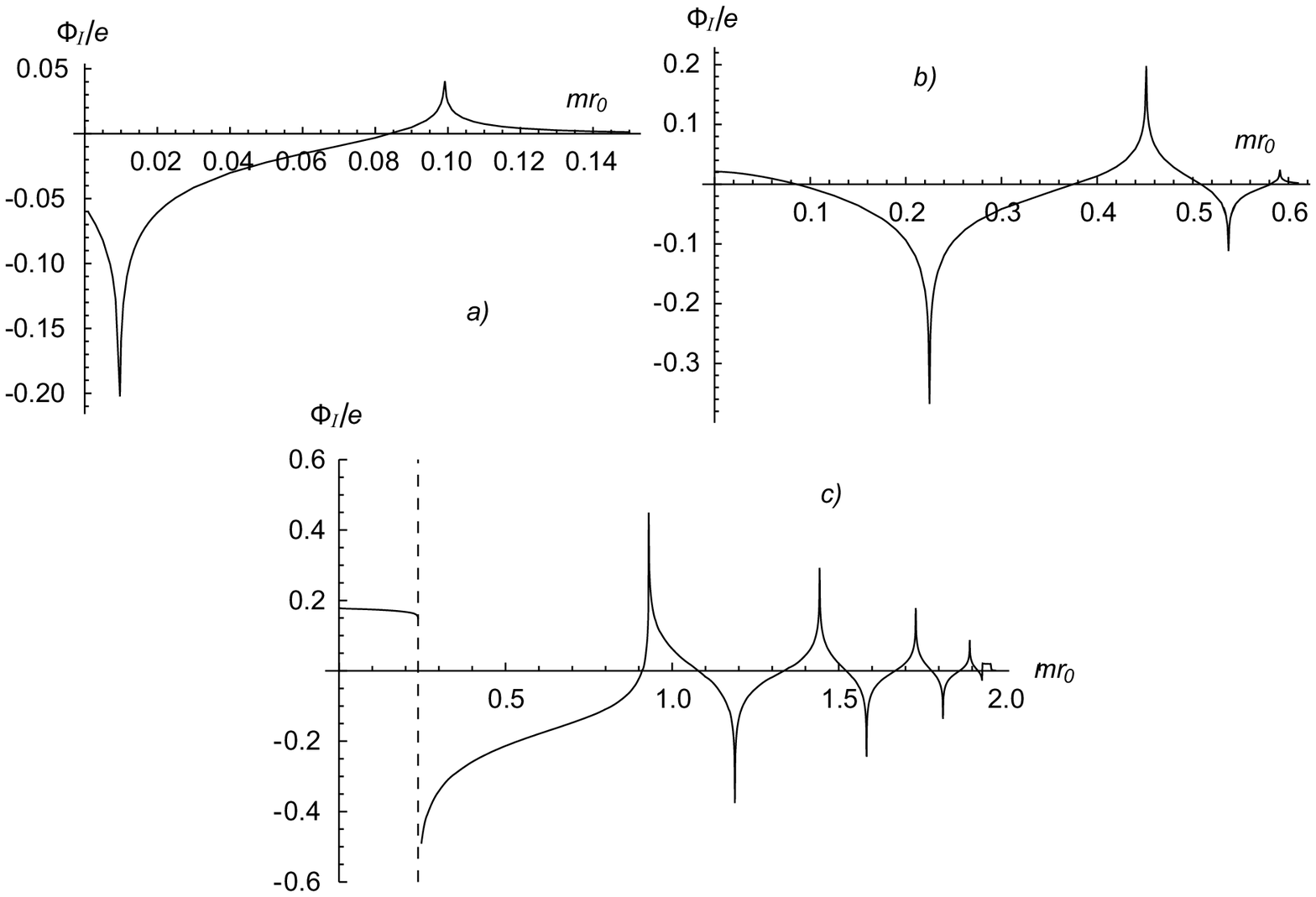}
    \caption{Dimensionless vacuum flux at $0 < \cot\theta < \infty$, $F=F_+$, $d=3$, and $\nu=1/2$: a) $\theta=3\pi/8$, b) $\theta=\pi/4$, c) $\theta=\pi/8$.}
    \label{fig:9}
\end{figure}

\newpage

\newpage

\begin{figure}[t]
    \centering
    \includegraphics[width=\textwidth]{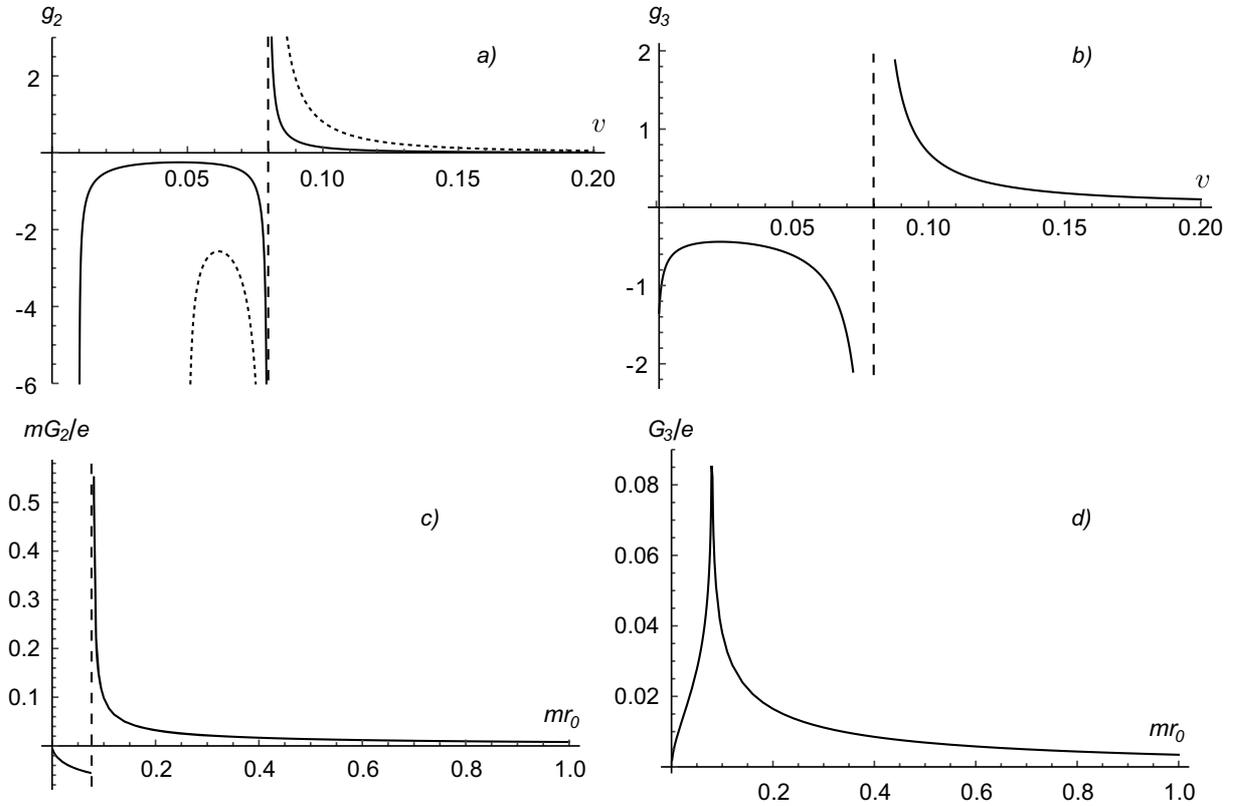}
    \caption{Eqs. \eqref{78} and \eqref{77} in the $d=2$ and $d=3$ cases: a) $g_2(v)$, solid and dotted lines correspond to $mr_0=10^{-2}$ and $mr_0=5\cdot10^{-2}$, b) $g_3(v)$, c) $m G_2/e$, d) $G_3/e$.}
    \label{fig:FirstTerm}
\end{figure}

\end{document}